\documentclass{jfm}

\usepackage[dvipsnames]{xcolor}
\usepackage{graphicx}
\usepackage{newtxtext}
\usepackage{newtxmath}
\usepackage[export]{adjustbox}
\usepackage{natbib}
\usepackage{tikz}
\usepackage{comment}
\usepackage{pdflscape}
\usepackage[hidelinks]{hyperref}

\newcommand{\RomanNumeralCaps}[1]

\title{\vspace{-20mm}
{ Coherence does not always imply causality in wall-bounded turbulence}}

\author{
  Peng E S Chen \aff{1}\corresp{\email{peng.c@upm.es}},
  \and
  Javier Jimenez \aff{1}
}
\affiliation{\aff{1}School of Aeronautics, Universidad Polit{\'e}nica de Madrid, 28040 Madrid, Spain.
}
\begin{document}
\maketitle
\begin{abstract}
Compact structures of intense tangential Reynolds stress (Q events) are well-known components of wall-bounded turbulence, and have been shown to be coherent because they approximately govern their own evolution. It has therefore often been assumed that they also are causally significant, in the sense that they explain the evolution of the incoherent component of the flow, which could thus be modeled as a superposition of structures. Since strong events typically only fill a small percentage of the total flow volume, this is also cited as a reason for considering structures  as targets for efficient engineering flow control. This paper shows that the causality assumption does not hold in general. Only about half of the structures identified in the flow are causally more significant than an equivalent volume of incoherent turbulence. To explain this variability, feature-based analysis and conditional averaging are performed. For wall-attached Q2 events and wall-detached structures, causally enhanced events are characterized by elevated strain rate and spanwise vorticity. These signatures are traced to intense upstream strain regions generated by the interaction of Q4-like motions impinging on Q2 events. For wall-attached Q4 events, the dominant indicators are instead the wall-normal and streamwise vorticity components; enhanced causal significance is associated with strong wall-normal vorticity. These findings show that quadrant events cannot be treated as a dynamically homogeneous class in causal analyses. Their causal significance depends strongly on the local flow environment, emphasizing the need to interpret coherent structures in terms of their interactions with the surrounding turbulence.
\end{abstract}

\section{Introduction}

Turbulent wall-bounded flows are characterized by coherent motions that play a central role in momentum transfer and energy redistribution. Among these, quadrant events, defined by the streamwise and wall-normal velocity fluctuations, have been extensively studied since the seminal works of \citet{wallace1972wall} and \citet{lu1973measurements}. Ejections (Q2) and sweeps (Q4), in particular, are known to make dominant contributions to the Reynolds shear stress \citep{wallace2016quadrant, lozano2012three} and are widely regarded as key elements of the near-wall turbulence cycle \citep{zhou1999mechanisms, jimenez2022streaks}. Recent studies of causality in wall-bounded turbulence, however, have shown that the causal significance of turbulent regions is highly non-uniform. Among randomly distributed perturbation regions, causally significant events have been found to occur preferentially near { negative wall-normal velocity regions, as} in Q4, whereas causally insignificant events are more commonly associated with { positive wall-normal velocity regions, as} in Q2 \citep{osawa2024causal}. These observations raise several questions. Are Q4 events, as a class, more causally significant than the surrounding turbulence? Are Q2 events generally less significant?
The present study addresses these questions.

To clarify the motivation for the present work, we first review previous approaches to the analysis of quadrant events, with particular attention to their limitations from a causal perspective. These approaches can be broadly divided into three categories. The first is classical quadrant analysis, in which quantities such as the Reynolds shear stress are conditioned on (the signs of) the velocity fluctuations \citep{willmarth1972structure}. Strictly speaking, this approach does not identify coherent structures, because no spatial or temporal connectivity is imposed \citep{robinson1991coherent}. It is therefore more appropriately described as quadrant analysis than as quadrant-event analysis.
The second category identifies intense quadrant events as spatially connected three-dimensional structures \citep{lozano2012three, lozano2014time, dong2017coherent, jimenez2018coherent}. This framework has enabled detailed investigations of their size, shape, spatial distribution and lifetime. However, such structures are often treated as isolated entities, which makes it difficult to determine their dynamical role in the evolution of turbulence. Conditional averages can reveal local associations between quadrant events and neighbouring structures, but they provide limited information about the global organization of the flow, and the inferred mechanisms often rely on phenomenological interpretation \citep{lozano2012three}.
The third category does not identify quadrant events explicitly. Instead, quadrant-like motions emerge when the velocity field is conditioned on other coherent structures \citep{zhou1999mechanisms, adrian2000vortex, liu2026conditionally}. Because the quadrant motions are not prescribed \textit{a priori}, the resulting associations are generally more robust. Nevertheless, as in the second category, these approaches remain primarily correlational and do not directly establish the causal role of quadrant events.

Recent developments in causal inference for dynamical systems provide a route to address this limitation. Two broad classes of approaches can be distinguished: observational and interventional. Observational approaches analyse existing data without perturbing the system. Temporal directionality can be used to infer possible cause--effect relationships, as in transfer-entropy analyses of turbulent flows \citep{lozano2020causality, martinez2024decomposing, ling2024causality},
{
Perron--Frobenius analysis \citep{page2020searching, jimenez2023perron},}
and related game-theoretic approaches \citep{cremades2024identifying}. However, observational methods are intrinsically limited by hidden confounding variables and 
cannot, by themselves, establish causation unequivocally. More generally, data-driven causal-inference methods widely used in social science, epidemiology and environmental science often rely on prescribed structural models or directed acyclic graphs to encode assumed dependencies between variables \citep{pearl2009causal, runge2023causal}. Although such frameworks provide a rigorous mathematical language for causal reasoning, they require an explicit representation of the relevant dependencies \citep{lozano2021cause}. In turbulence, constructing such a representation for interacting coherent structures is challenging because of the high dimensionality, strong nonlinearity and continuous spatio-temporal nature of the flow \citep{cardesa2017turbulent}. Purely observational approaches therefore face intrinsic limitations in establishing causal structure in turbulence.

Interventional approaches, by contrast, deliberately perturb selected components of the system and measure the resulting response. Advances in high-performance computing have made such interventions feasible in direct numerical simulations. \citet{jimenez2018machine} introduced localized perturbations in two-dimensional turbulence and identified vortex dipoles as dynamically influential structures. \citet{encinar2023identifying} extended this methodology to three-dimensional homogeneous isotropic turbulence and showed that significant perturbations tend to align with the most stretching eigenvector of the strain tensor. More recently, \citet{osawa2024causal} applied a related framework to open-channel turbulence and found that significant perturbations are preferentially advected towards the wall, suggesting a connection between causal influence and sweep-like motions. Although such methods have begun to reveal dynamically influential flow regions, their use in quantifying the causal role of specific coherent structures remains limited.
{ What remains unclear is whether structures identified by classical coherence criteria are also causally significant in an interventional sense.}

Closely related ideas have also been explored in the context of predictability. Optimization-based approaches identify perturbations that maximize energy growth over finite time horizons \citep{cherubini2010optimal, farano2017optimal, ciola2023nonlinear}, thereby highlighting dynamically sensitive flow configurations. However, these methods are usually restricted to relatively short time horizons and can depend strongly on the particular flow realization. More recently, information-theoretic frameworks, including those developed by Avila and co-workers, have used measures such as the Kullback--Leibler divergence to quantify the predictability of extreme events \citep{vela2024large, montesdeoca2025probabilistic}. These approaches provide valuable insight into flow sensitivity and predictability, but they do not directly quantify causal influence in the interventional sense considered here.

In this paper, causal significance is used in an explicitly interventional sense: a flow region is regarded as causally significant if a localized perturbation applied to that region produces, at a given subsequent observing time, a larger dynamical response than an otherwise comparable perturbation applied elsewhere.
We use pairwise interventional experiments to assess the causal significance of quadrant events in turbulent channel flow at moderate Reynolds number. The treated group consists of quadrant structures, while the corresponding comparison group is constructed from flow regions with the same geometry and wall-normal location but displaced in the wall-parallel directions. Since quadrant events occupy less than $10\%$ of the total flow volume \citep{lozano2012three}, the comparison group predominantly samples the background turbulence. This construction separates the effect of coherent dynamics from the purely geometrical contribution associated with the perturbation region. Causal significance is quantified by the perturbation energy generated by localized disturbances introduced within the structures at the initial time \citep{osawa2024causal}. The causal effect associated with coherence is then measured by comparing the perturbation energy of quadrant events with that of their paired comparison regions.

The results show that quadrant events are not uniformly more causally significant than the background turbulence. A substantial fraction of both Q2 and Q4 events exhibits lower causal significance than their paired comparison regions. To identify the origin of this variability, we perform feature-based analysis and conditional averaging. The results show that the causal significance of quadrant events depends strongly on their local dynamical environment and, in particular, on their interactions with neighbouring structures. These findings indicate that quadrant events should not be treated as a dynamically homogeneous class in causal analyses. Rather, their causal importance is contingent on the surrounding flow field.

The remainder of the paper is organized as follows. Section~\ref{sec:methodology} describes the methodology, including the design of the pairwise interventional experiments. The results are presented in \S\ref{sec:results}. Finally, \S\ref{sec:conclusions} summarizes the main findings and discusses their implications.

\section{Pairwise interventional experiments}
\label{sec:methodology}

The interventional procedure is designed to compare the dynamical response induced by perturbing quadrant structures with that induced by perturbing geometrically identical regions of background turbulence. Statistically independent velocity fields, hereafter referred to as base flows, are first collected under identical flow conditions (as shown in Table \ref{tab:setup}), and quadrant structures are identified in each snapshot. For each structure, a paired comparison region is constructed by shifting the same geometrical mask to a randomly selected position in the streamwise and spanwise directions, while preserving its wall-normal location and shape. The original quadrant structures are referred to as non-shifted structures, whereas their displaced counterparts are referred to as shifted structures.
The velocity field within each non-shifted or shifted structure is then modified to construct a perturbed initial condition. The unmodified and modified fields are evolved separately, yielding the reference flow $\boldsymbol{u}_r$ and the modified flow $\boldsymbol{u}_m$, respectively. The resulting perturbations are used to quantify the causal significance of the perturbed region, as described below.



\begin{table}
\centering
{\small
\begin{tabular}{cccccccc}
 $Re_{\tau}$ & $\Delta^{+}$ & $(L_x^+, L_z^+)$ 
& $\Delta t^+$ & $(N_x, N_y, N_z)$ 
& $(\Delta x^{+}, \Delta y_w^{+}, \Delta z^{+})$ 
& $E/K$ & $N_{qs}$ \\
 600 
& 79 
& $(1884, 1884)$ 
& $2.15\times 10^{-4}$ 
& $(192, 393, 394)$ 
& $(14, 0.46, 7.2)$ 
& $1.77\times10^{-4}$ 
& 17149 \\
\end{tabular}
}
\caption{Interventional experiment setups. See text for detailed definitions.}
\label{tab:setup}
\end{table}

\subsection{Base flow}

We consider incompressible turbulent channel flow at moderate Reynolds number. The governing equations are the incompressible Navier--Stokes equations,
\begin{equation}
    \partial_t \boldsymbol{u}^{o}
    + \boldsymbol{u}^{o} \cdot \nabla \boldsymbol{u}^{o}
    =
    -\nabla p^{o}
    + \frac{1}{Re}\Delta \boldsymbol{u}^{o},
    \label{eq:NS}
\end{equation}
\begin{equation}
    \nabla \cdot \boldsymbol{u}^{o}=0,
    \label{eq:incompressibility}
\end{equation}
where $\boldsymbol{u}^{o}=(u^{o},v^{o},w^{o})$ is the velocity field in streamwise ($x$), wall-normal ($y$) and spanwise ($z$) directions, $p^{o}$ is the pressure and the density is set to unity. The bulk velocity ($U_b$) and half channel height ($h$) are used to normalize the equation. The Reynolds number is defined as $Re=U_b h/\nu$, where $\nu$ is the kinematic viscosity. {Quantities with a superscript $+$ are normalized with wall units, i.e. $u_\tau$ and $\nu$, where $u_\tau$ is the friction velocity.} The velocity components are decomposed into mean and fluctuating parts as $u^{o}=U+u, v^{o}=V+v, w^{o}=W+w $, { where uppercase letters denote mean quantities and lowercase letters denote fluctuations}.
The equations are solved by direct numerical simulation. The numerical parameters are summarized in Table~\ref{tab:setup}. The friction Reynolds number is $Re_{\tau}=u_\tau h/\nu=600$, which is sufficiently high to reduce low-Reynolds-number effects while allowing a large number of independent interventional experiments. The computational domain is $L_x/h=L_z/h=\pi$, corresponding to approximately $1884$ wall units in both homogeneous directions. This domain size is sufficient to obtain reliable turbulence statistics below $y^+\approx 400$, while also containing a substantial number of large quadrant structures \citep{flores2010hierarchy, lozano2014effect}.
In order to march the modified velocity and the reference velocity easily, the time spacing is set to be constant as $\Delta t^+\approx 2.15\times 10^{-4}$.

Spatial discretization is performed using dealiased Fourier spectral methods in the homogeneous directions \citep{kim1987turbulence} and seven-points-stencil compact finite differences in the wall-normal direction \citep{hoyas2006scaling}. Time advancement uses a semi-implicit third-order Runge--Kutta scheme \citep{spalart1991spectral}. The mass flux is held constant, and the wall-normal grid is stretched using a hyperbolic-tangent distribution. The code has been ported to GPU architectures \citep{vela2021entropy, osawa2024causal}.

\begin{figure}
    \centering
    \includegraphics[width=\linewidth]{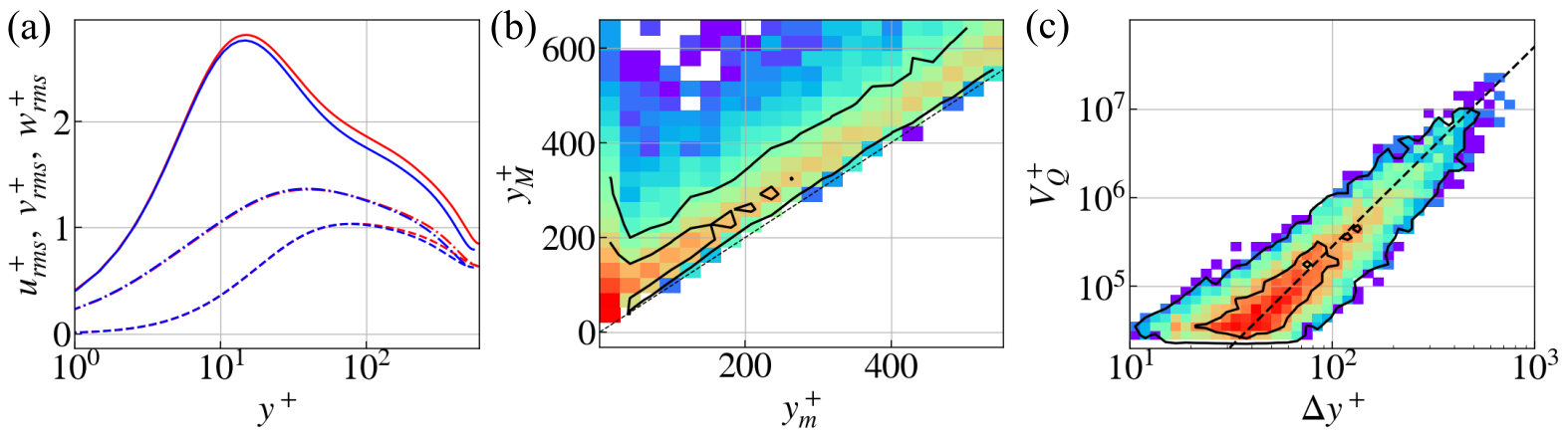}
    \caption{(a) Turbulence intensity of the reference velocity field: solid line $u^+_{rms}$, dashed lines $v^+_{rms}$, dash-dotted lines $w^+_{rms}$. The red lines are our data with $Re_\tau=600$, while the blue ones are those of \citet{del2003spectra} with $Re_\tau=550$.(b) Joint probability density function (JPDF) of the minimum and the maximum wall-normal position of the structures. (c) JPDF of the volume and the height of the wall-attached Qs. The black solid lines contain 50\% and 98\% data. The black dashed line is $V_Q^+\sim (\Delta y^+)^{2.25}$. }
    \label{fig:urms-vol-ymin}
\end{figure}

A total of 40 statistically independent snapshots are collected after the flow reaches a statistically stationary state, with consecutive snapshots separated by over 1.5 eddy-turnover times ($h/u_\tau$). Figure~\ref{fig:urms-vol-ymin}(a) shows the turbulence intensities. The one-point statistics are in good agreement with those reported in the literature at comparable Reynolds numbers \citep{del2003spectra}.

\begin{table}
    \centering
    \begin{tabular}{cccccccc}
      Type &    $V$ &  $V_{att}$ &  $V_{det}$ &    $N$ &  $N_{att}$ &  $N_{det}$ \\
      Q1 & 0.0026 &     0.0002 &     0.0024 & 0.1470 &     0.0103 &     0.1367 \\
      Q2 & 0.0134 &     0.0060 &     0.0073 & 0.3535 &     0.1532 &     0.2002 \\
      Q3 & 0.0024 &     0.0001 &     0.0023 & 0.1394 &     0.0029 &     0.1365 \\
      Q4 & 0.0150 &     0.0065 &     0.0085 & 0.3601 &     0.1366 &     0.2236 \\
    
    \end{tabular}
    \caption{Volume fraction with respect to the total volume of the channel and number fraction with respect to the total number of the quadrant structures. The subscript $att$ denotes the wall-attached structures, while $det$ denotes the wall-detached structures.}
    \label{tab:vol-num}
\end{table}

\subsection{Quadrant-event identification}

Following \citet{lozano2012three}, quadrant events are identified as connected regions satisfying
\begin{equation}
    |uv| \ge H u_{\mathrm{rms}}v_{\mathrm{rms}},
    \label{eq:qs-definition}
\end{equation}
where $H=1.75$. Each connected region is classified according to the signs of the volume-averaged velocity fluctuations,
\begin{equation}
    u_Q =
    \frac{1}{V_Q}\int_Q u\,\mathrm{d}V,
    \qquad
    v_Q =
    \frac{1}{V_Q}\int_Q v\,\mathrm{d}V,
\end{equation}
where $Q$ denotes the occupied region of the structure and $V_Q$ is the volume of the structure.
The quadrant type is then defined as
\begin{equation}
Q =
\begin{cases}
\mathrm{Q1}, & u_Q>0,\ v_Q>0, \\
\mathrm{Q2}, & u_Q<0,\ v_Q>0, \\
\mathrm{Q3}, & u_Q<0,\ v_Q<0, \\
\mathrm{Q4}, & u_Q>0,\ v_Q<0 .
\end{cases}
\label{eq:quadrant-classification}
\end{equation}
Structures with volume $V_Q^+<30^3$ are excluded. Structures with streamwise or spanwise extent larger than half the corresponding domain length, $\Delta x>L_x/2$ or $\Delta z>L_z/2$, are also removed to avoid finite-domain effects.

Connected regions are identified using the density-based spatial clustering algorithm DBSCAN \citep{ester1996density}, which groups points according to their local spatial connectivity and is controlled by two parameters: the neighbourhood radius $\varepsilon$ and the minimum number of points required to form a dense region, $N_{\min}$. In the present study, these parameters are set to $\varepsilon=\sqrt{3}$ in grid-index space and $N_{\min}=8$. With this choice, the resulting segmentation is equivalent to the connectivity criterion used by \citet{moisy2004geometry}. The Python implementation of DBSCAN is used for convenience and computational efficiency.

{
A total of 17149 structures are identified. 
Figure~\ref{fig:urms-vol-ymin}(b) shows the joint probability density function of $y_m^+$ and $y_M^+$, where $y_m^+$ and $y_M^+$ denote the minimum and maximum wall distances of the structure, respectively. Two distinct branches are observed, motivating the classification of structures as wall-attached and wall-detached.
Following \citet{del2006self} and \citet{lozano2012three}, wall-attached structures are those with $y_m^+<20$, while wall-detached structures are those with $y_m^+\ge20$.
Figure~\ref{fig:urms-vol-ymin}(c) shows the joint probability density function of the volume $V_Q^+$ and wall-normal height $\Delta y^+=y_M^+-y_m^+$ for wall-attached quadrant structures. The distribution follows the scaling
\[
    V_Q^+ \sim (\Delta y^+)^{2.25},
\]
in agreement with \citet{lozano2012three}.
Table~\ref{tab:vol-num} lists the volume fractions and number fractions of the quadrant structures. 
The volume fractions are smaller than those reported by \citet{lozano2012three}, mainly because very large structures are removed in the present analysis. Such structures cannot be fully accommodated within the small computational domain and are therefore excluded to avoid domain-size effects.
These results indicate that the structures identified in the present simulations are consistent with those studied by \citet{lozano2012three}. 
Since wall-attached Q1 and Q3 structures are rare and contribute only weakly to the total population, the following analysis focuses on wall-attached Q2 and Q4, and wall-detached Q1, Q2, Q3, and Q4.
}

\begin{figure}
    \centering
    \includegraphics[width=\textwidth]{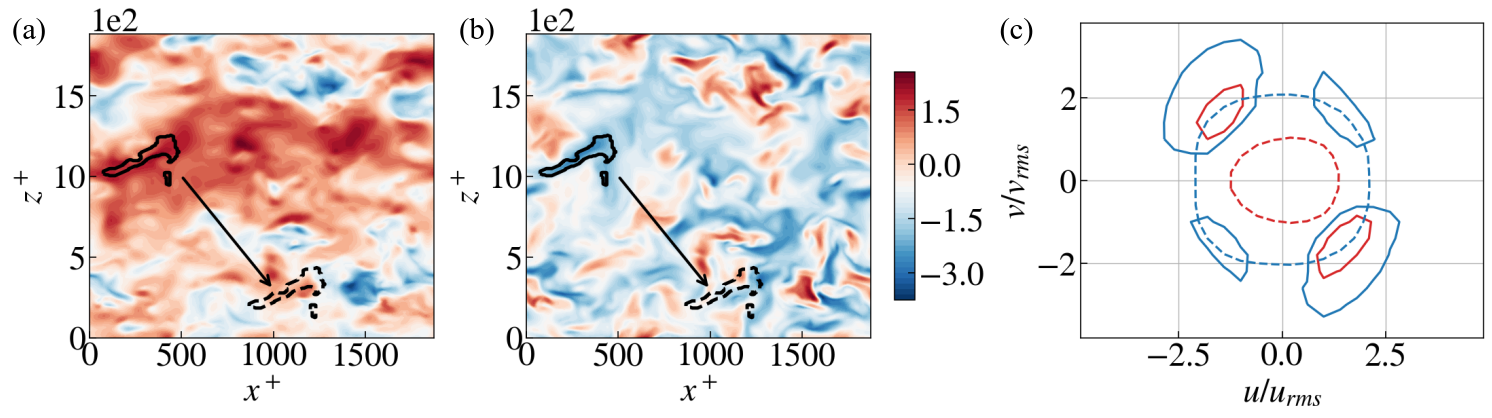}
    \caption{(a,b) Sketch of the structure shifting. The background color contours show the streamwise velocity fluctuation in (a) and the wall-normal velocity fluctuation in (b). The original structure is enclosed by the solid lines, while the shifted structure is enclosed by the dashed lines. (c) Joint probability density function (JPDF) of $u$ and $v$ within the non-shifted structures (solid lines) and the shifted structures (dashed lines) . The contours enclosing $50\%$ and $90\%$ of the probability mass are shown in blue and red, respectively. }
    \label{fig:uv-shifted}
\end{figure}

The structures are then shifted in the $x$ and $z$ directions while preserving their geometry and wall-normal position, as illustrated in figure~\ref{fig:uv-shifted}(a,b). After this displacement, the shifted mask is no longer conditioned on the original velocity-fluctuation field. Consequently, the criterion \eqref{eq:qs-definition} and the associated quadrant signs \eqref{eq:quadrant-classification} are generally not satisfied within the shifted region, as shown in figure~\ref{fig:uv-shifted}(c). By contrast, geometrical quantities such as the volume, $y_m^+$ and $y_M^+$ are unchanged. The shifted structures therefore provide geometrically matched comparison regions that predominantly sample the background turbulence.
No additional exclusion criterion is imposed when a shifted region overlaps another intense event. Such cases are retained and treated as part of the background sampling. In practice, these overlaps are rare because quadrant structures occupy only a small fraction of the total flow volume.

\subsection{Initial perturbation}

For each non-shifted or shifted structure, the velocity field is modified locally to construct a perturbed initial condition. The modified velocity field is defined as
\begin{equation}
    \boldsymbol{u}_m
    =
    \boldsymbol{u}_r(1-\beta g)
    +
    \left(\boldsymbol{u}_s-\boldsymbol{\alpha}\right)\beta g,\text{  at } t=0, 
    \label{eq:modified-field}
\end{equation}
where $\boldsymbol{u}_r$ is the reference velocity field and $\boldsymbol{u}_s$ is a Gaussian-filtered version of $\boldsymbol{u}_r$. The filter width is set to $\Delta^+=79$, which is larger than the smallest retained structures and comparable to the perturbation box width used by \citet{osawa2024causal}. The function $g$ is a smooth blending mask that is approximately unity inside the structure and decays smoothly to zero outside it.
Because the quadrant structures have complex three-dimensional shapes, the construction of $g$ requires some care. At fixed $y$ and $z$, the characteristic function of a structure may consist of several disconnected intervals in the streamwise direction. Let these intervals have centres $x_{c,i}$ and widths $\delta_i$. For each interval, we define a smooth one-dimensional mask
\begin{equation}
    g_{x,i}
    =
    \frac{1}{2}
    \left[
    \tanh\left(
    \frac{\delta_i-|x-x_{c,i}|}{\delta_0}
    \right)
    +1
    \right].
    \label{eq:gx}
\end{equation}
where $\delta_0=1000 h$ is a constant.
The streamwise mask is then obtained as
\begin{equation}
    g_x = \sum_i g_{x,i}.
\end{equation}
The wall-normal and spanwise masks, $g_y$ and $g_z$, are constructed analogously, and the full three-dimensional mask is defined as
\begin{equation}
    g = g_x g_y g_z .
\end{equation}

The correction term
\begin{equation}
\boldsymbol{\alpha}(y)
=
\frac{
\displaystyle
\int g\left(\boldsymbol{u}_{s}-\boldsymbol{u}_{r}\right)
\,\mathrm{d}x\,\mathrm{d}z
}{
\displaystyle
\int g\,\mathrm{d}x\,\mathrm{d}z
},
\label{eq:alpha}
\end{equation}
is introduced to preserve the mean velocity profile at each wall-normal location. The normalization factor
\begin{equation}
\beta
=
\left[
\frac{
E L_x L_y L_z
}{
\displaystyle
\int g
\left|
\boldsymbol{u}_{s}
-\boldsymbol{\alpha}
-\boldsymbol{u}_{r}
\right|^{2}
\,\mathrm{d}V
}
\right]^{1/2},
\label{eq:beta}
\end{equation}
sets the prescribed total perturbation energy to $E/K=1.77\times10^{-4}$, where $K$ is the turbulent kinetic energy of the base flow. This energy level is comparable to that used by \citet{osawa2024causal}. Since the volumes of the structures differ, imposing a fixed total perturbation energy leads to different mean perturbation energies within individual structures. This choice is intentional: it mimics a control setting in which the total input energy is fixed. Finally, a pressure-projection step is applied to enforce incompressibility of the modified velocity field \citep{encinar2023identifying}.
The pressure-projection step modifies the total perturbation energy, and its influence is discussed in Appendix \ref{app:press}.

For each interventional experiment, corresponding to either a shifted or a non-shifted structure, the simulation is advanced to $t=0.5h/u_{\tau}$. Flow fields are sampled every $0.01h/u_{\tau}$, yielding 50 snapshots per experiment. In total, the simulations required approximately 640 GPU-days on NVIDIA A100 GPUs. 
{
If all instantaneous fields were stored, the resulting data set would be approximately 1 PB in size. To avoid archiving such a large data set, the post-processing was performed on the fly, and only the required statistics were stored.
}

\subsection{Quantification of coherence effects}
The responding perturbation energy is defined as the volume-averaged perturbation energy over the entire channel,
\begin{equation}
    k = \frac{1}{V_{ch}} \int_{ch} \|\boldsymbol{u}_m - \boldsymbol{u}_r\|^2 \, \mathrm{d}V,
\end{equation}
where $V_{ch}=L_xL_yL_z$ is the total volume of the channel.
Thus, although the initial modification is localized within a quadrant structure or its shifted counterpart, the response is measured globally. This definition allows $k(t)$ to quantify the total dynamical effect of the localized intervention on the flow. 
Note that $k(0)\ne E$ because of the pressure step, see Appendix \ref{app:press} for details.
Other norms could be considered, but the $L^2$ norm provides a natural and robust choice.
{
The governing equation of $k$ can be derived from the governing equations for $\boldsymbol{u}_m$ and $\boldsymbol{u}_r$ \citep{nikitin2018characteristics, osawa2024causal}, 
\begin{equation}
    \partial_t k = P-\varepsilon.
\end{equation}
Here,
\begin{equation}
    P = -\frac{1}{V_{ch}} \int_{ch} (u_{i,m}-u_{i,r}) (u_{j,m}-u_{j,r}) S_{ij, r} dV,
    \label{eq:prod}
\end{equation}\begin{equation}
    \varepsilon = \frac{2\nu}{V_{ch}} \int_{ch}  (S_{ij, m} - S_{ij, r}) (S_{ij, m} - S_{ij, r})dV,
    \label{eq:diss}
\end{equation}
and $S_{ij,r} = \frac{1}{2}(\partial_i u_{j,r}+\partial_j u_{i,r})$ and $S_{ij,m} = \frac{1}{2}(\partial_i u_{j,m}+\partial_j u_{i,m})$ are the strain tensors of the reference and modified flows, respectively.
}
The effect of quadrant coherence is quantified by comparing the perturbation response of each non-shifted structure with that of its shifted counterpart. For a given pair, we define the enhancing factor
\begin{equation}
    F
    =
    \ln
    \frac{k_{\mathrm{ns}}(t)}
         {k_{\mathrm{s}}(t)},
    \label{eq:F}
\end{equation}
where $k_{\mathrm{ns}}(t)$ and $k_{\mathrm{s}}(t)$ are the perturbation energies at the target time $t$ for the non-shifted and shifted structures, respectively. 
Positive values of $F$ indicate that the original quadrant structure produces a stronger perturbation response than its geometrically matched shifted counterpart. Negative values indicate that the same geometry is more influential when placed in the background flow than when placed on the original quadrant event. Thus, $F$ measures the causal effect associated with the original quadrant-based velocity organization, after removing the contribution of geometry and wall-normal position.
For conditional analysis, structures are classified according to $F$. The top and bottom $10\%$ of the distribution at the most distinguished time (see \S \ref{sec:time-evolution} for detailed definition) are referred to as causally enhanced and causally suppressed events, respectively, while the central portion of the distribution is treated as neutral.

\section{Results}
\label{sec:results}

\subsection{Time evolution of enhancing factor}
\label{sec:time-evolution}
{
\begin{figure}
    \centering
    \includegraphics[width=\linewidth]{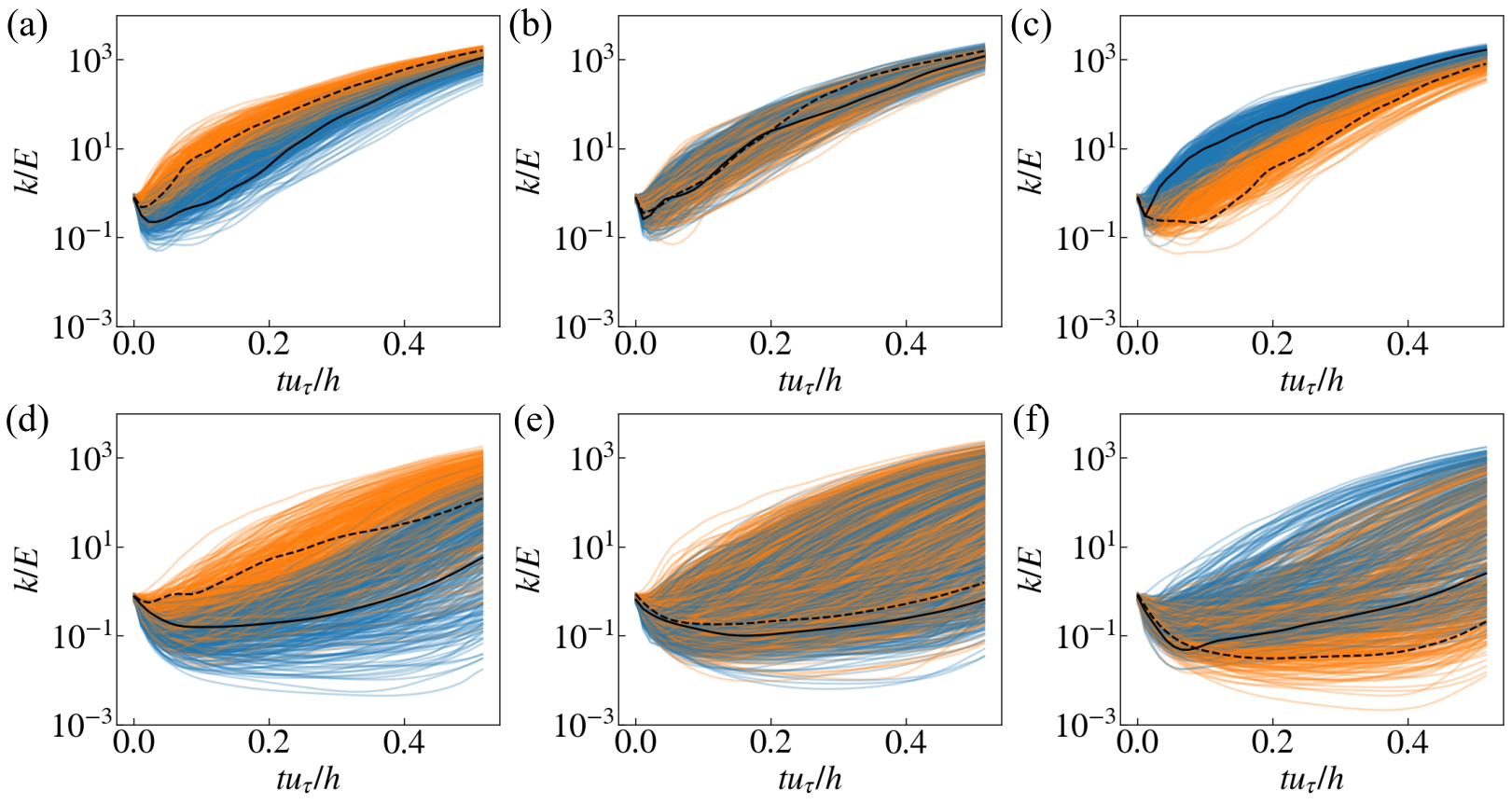}
    \caption{Time evolution of the perturbation energy for (a,b,c) wall-attached Q2 events and (d,e,f) wall-detached events. 
    (a,d) Causally suppressed events; (b,e) neutral events; (c,f) causally enhanced events. 
    Blue lines denote non-shifted structures, and yellow lines denote their shifted counterparts. 
    One randomly chosen representative pair is highlighted in black: the solid line corresponds to the non-shifted structure, whereas the dashed line corresponds to its shifted counterpart.}
    \label{fig:k-att-det-q2}
\end{figure}
Figure~\ref{fig:k-att-det-q2} shows the time evolution of the total perturbation energy for wall-attached Q2 and wall-detached Q2. The other classes of quadrant events exhibit qualitatively similar behaviour. In all cases, the perturbation energy initially decreases before undergoing rapid growth, which remains sub-exponential, as in \citet{osawa2024causal}. For causally suppressed events, the perturbation energy generated by the non-shifted structures is lower than that generated by their shifted counterparts. Conversely, for causally enhanced events, the non-shifted structures produce a stronger perturbation response than their shifted counterparts.
Compared with wall-attached Q2, the perturbation energy of wall-detached Q2 is weaker at all times. This is mainly because wall-detached Q2 has a longer decreasing interval.
}
\begin{figure}
    \centering
    \includegraphics[width=\linewidth]{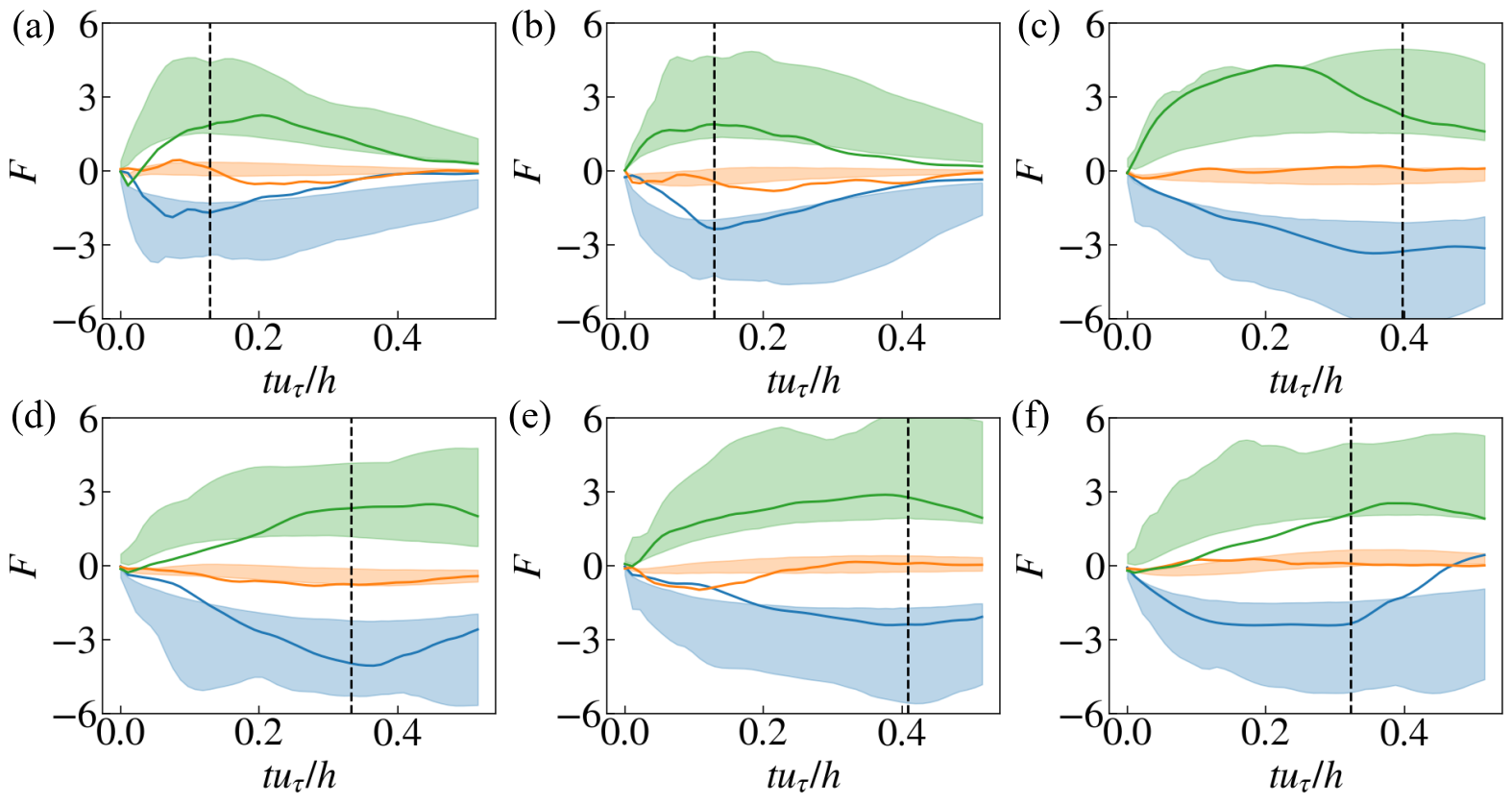}
    \caption{Time evolution of the enhancing factor $F$ for (a) wall-attached Q2, (b) wall-attached Q4, (c) wall-detached Q1, (d) wall-detached Q2, (e) wall-detached Q3, and (f) wall-detached Q4. The shaded regions correspond to the top 10\% (green), the central 40--60\% (yellow), and the bottom 10\% (blue) ranges of $F$, computed independently at each observation time. Representative trajectories from each region are shown as coloured solid lines. The dashed black lines stand for the maximum separation between the 90th percentile (lower bound of the green region) and the 10th percentile (upper bound of the blue region).}
    \label{fig:F-qs}
\end{figure}

Figure \ref{fig:F-qs} presents the time evolution of the enhancing factor $F$.
Three characteristic regions are highlighted: the top 10\% (green), the central 40--60\% (yellow), and the bottom 10\% (blue).
{ In this figure, these percentile ranges are recomputed independently at each observation time in order to illustrate the evolution of the distribution of $F$. By contrast, the causally enhanced and causally suppressed sets used in the subsequent feature and conditional analyses are defined at the distinguished time $t_d$, where the separation between the enhanced and suppressed regions is maximal.}
The first notable observation from figure \ref{fig:F-qs} is that the top and bottom 10\% regions are approximately symmetric about $F=0$, while the central region remains close to $F=0$. 
This near-symmetry suggests that the coherent dynamics do not preferentially enhance or suppress the causal significance of the structures, and that the net effect of the dynamics is approximately neutral. 
This observation indicates that not all but only approximately half of the quadrant events are more causally significant than the background turbulence. 
A second observation is that the magnitude $|F|$ initially increases with time, reaches a maximum, and subsequently decreases. 
{
For wall-attached events, the distinguished time $t_d= 0.13 h/u_\tau$.
For wall-detached events, $t_d\approx 0.3\sim 0.4 h/u_\tau$ is much later: $t_d=0.40h/u_\tau$ for wall-detached Q1, $t_d=0.33h/u_\tau$ for wall-detached Q2, $t_d=0.40h/u_\tau$ for wall-detached Q3, and $t_d=0.32h/u_\tau$ for wall-detached Q4.}
{The subsequent decrease of $|F|$ can be attributed to the nonlinear evolution and eventual saturation of the perturbation energy. If the perturbations grew linearly with a constant Lyapunov exponent, the ratio between the perturbation energies of a non-shifted structure and its shifted counterpart would remain constant, and hence $F$ would not vary in time for each pair. In the present simulations, however, the perturbation energy soon exceeds the linear-growth regime. 
{
At sufficiently long times, the modified and reference flows become decorrelated, so that both $k_{ns}(t)$ and $k_s(t)$ approach $2K$, where $K$ is the turbulent kinetic energy. As a result, $|F|$ approaches zero.
}
The distinguished time $t_d$ is substantially later for wall-detached events than for wall-attached events (as shown in figure \ref{fig:k-att-det-q2} (d,e,f)). This difference may be attributed to the weaker perturbation response of wall-detached events. Because their perturbation energy grows more slowly, the contrast between enhanced and suppressed responses takes longer to develop, and the maximum separation therefore occurs at a later time.
}

\begin{figure}
    \centering
    \includegraphics[width=\linewidth]{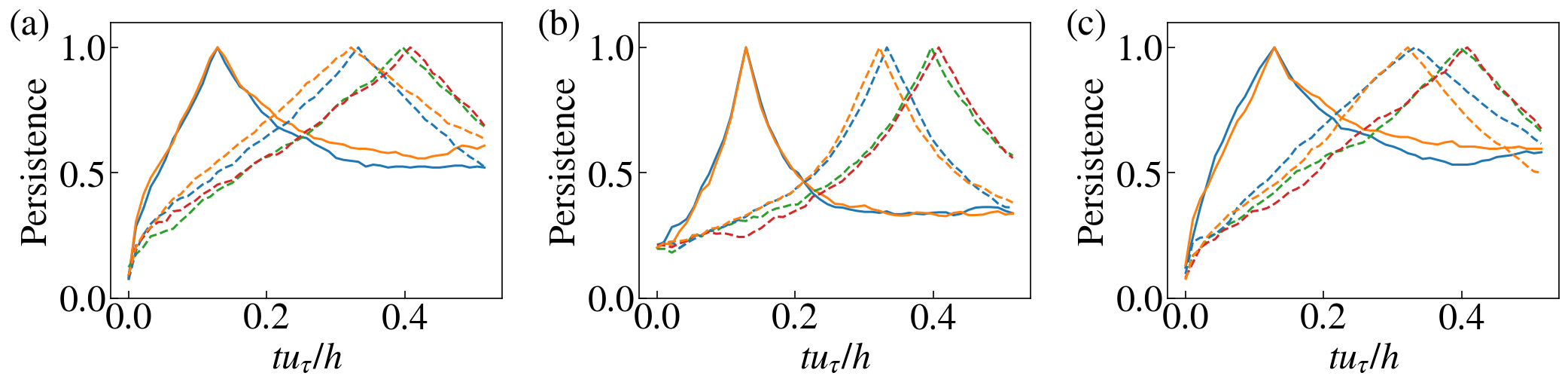}
    \caption{Persistence of (a) the most enhanced structures, (b) the null-effect structures, and (c) the most suppressed structures. The reference sets are defined at $t_d$. The lines denote: blue solid, wall-attached Q2; yellow solid, wall-attached Q4; blue dashed, wall-detached Q2; yellow dashed, wall-detached Q4; red dashed, wall-detached Q1; green dashed, wall-detached Q3.}
    \label{fig:persistence}
\end{figure}

The sample trajectories (solid lines in figure \ref{fig:F-qs}) indicate that individual realizations do not remain within a given percentile region at all times.
This behavior is related to the persistence of the percentile regions \citep{osawa2024causal}, which can be quantified as a conditional probability. For example, the persistence of the top 10\% region for wall-attached Q2 events is defined as
{
\begin{equation}
    P\left(\mathcal{T}_{Q2, att}^{F}(t) \mid \mathcal{T}_{Q2, att}^{F}(t_d)\right) = 
    \frac{P\left(\mathcal{T}_{Q2, att}^{F}(t)\cap \mathcal{T}_{Q2, att}^{F}(t_d)\right)}{P\left(\mathcal{T}_{Q2, att}^{F}(t_d)\right)},
\end{equation}
where $\mathcal{T}_{Q2, att}^{F}(t)$ denotes the top $10\%$ of wall-attached Q2 events ranked by $F(t)$.
}
Figure \ref{fig:persistence} shows the results. For wall-attached Q2 and Q4 events, the persistence exhibits similar behaviour: it increases initially, then decreases nonlinearly, and eventually saturates at approximately $0.5$ for the most enhanced region, $0.6$ for the most suppressed region, and $0.3$ for the central region.
For wall-detached structures, the increase and decrease are approximately linear, and such saturation is not observed within the available time window, possibly due to the limited duration of the simulations. In all cases, except at very early times, the persistence is significantly higher than the $10\%$ expected from a random selection.

\begin{figure}
    \centering
    \includegraphics[width=\linewidth]{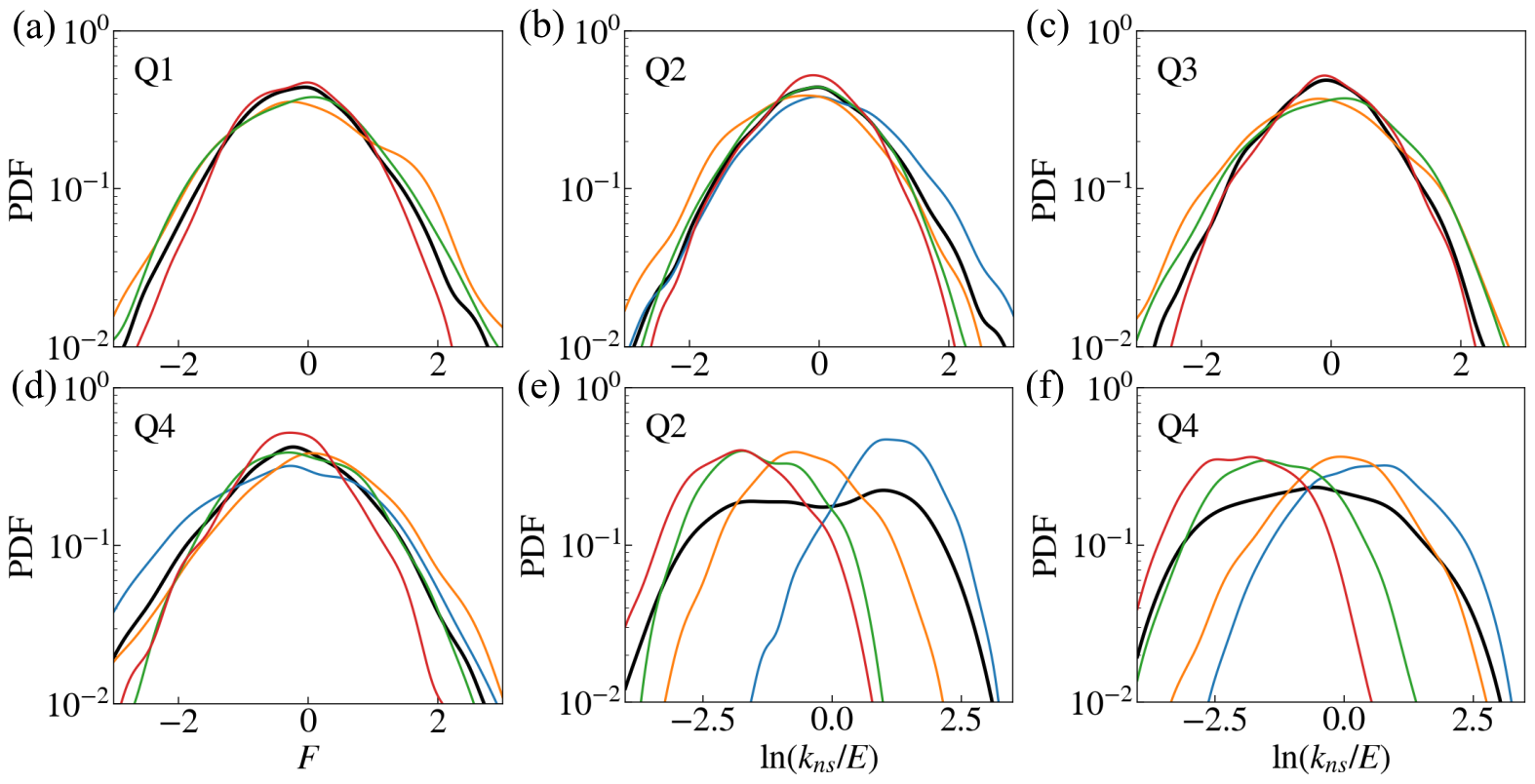}
    \caption{Probability density function of $F$ conditioned on $y_m^+$ at $t=0.1h/u_\tau$ for (a) Q1, (b) Q2, (c) Q3 and (d) Q4. The black solid lines denote the unconditional PDFs, while the coloured lines denote PDFs conditioned on $y_m^+$: blue, $y_m^+ \in [0,20)$; yellow, $y_m^+ \in [20,100)$; green, $y_m^+ \in [100,200)$; and red, $y_m^+ \in [200,400)$. (e,f) PDF of $\ln(k_{ns}/E)$ conditioned on $y_m^+$ at $t=0.1h/u_\tau$ for (e) Q2 and (f) Q4. Colours in (e,f) are the same as in (a-d).}
    \label{fig:jpdf-F-ymp}
\end{figure}

\begin{figure}
    \centering
    \includegraphics[width=0.66\linewidth]{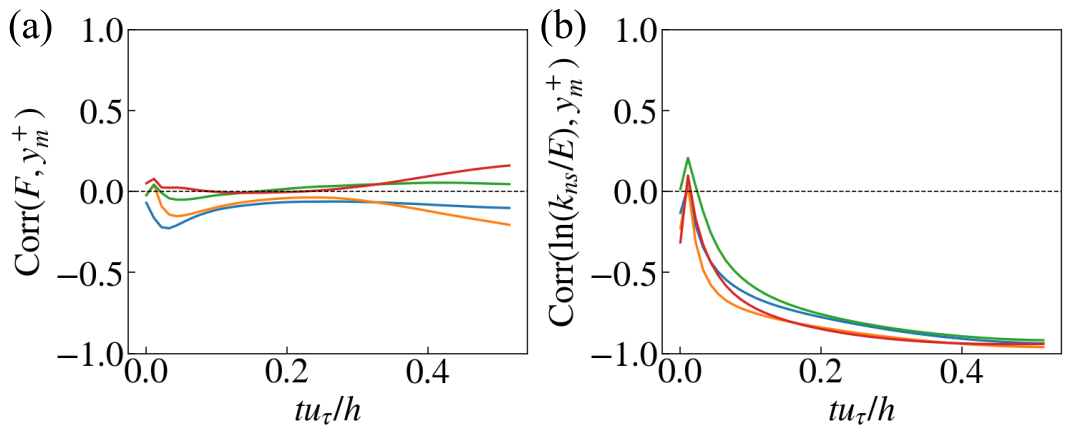}
    \caption{Correlation coefficients: (a) between $F$ and $y_m^+$, and (b) between $\ln (k_{ns}/E)$ and $y_m^+$. The coloured lines represent different quadrants: blue, Q1; red, Q2; yellow, Q3; green, Q4.}
    \label{fig:corr-F-ymp}
\end{figure}

Another notable result is that $F$ is only weakly correlated with the wall-normal location at which the initial perturbation is imposed, although the perturbation energy itself depends strongly on this location. 
{
Figure~\ref{fig:jpdf-F-ymp}(a--d) presents the probability density function of $F$ conditioned on $y_m^+$ at $t=0.1h/u_\tau$, i.e. $P(F\mid y_m^+)$. 
For comparison, figure~\ref{fig:jpdf-F-ymp}(e,f) shows the probability density function of $\ln(k_{ns}/E)$ conditioned on $y_m^+$ for Q2 and Q4 at the same time. Note that $y_m^+$ is independent of time, whereas both $F$ and $k$ evolve in time.
The behaviour at other times is qualitatively similar and is therefore omitted for brevity. For different values of $y_m^+$, the conditional distributions $P(F\mid y_m^+)$ are much closer to the unconditional distribution $P(F)$ than $P(\ln(k_{ns}/E)\mid y_m^+)$ is to $P(\ln(k_{ns}/E))$. This confirms that $F$ is only weakly dependent on $y_m^+$. 
}
This is further quantified in figure~\ref{fig:corr-F-ymp}(a), which shows that the correlation coefficient between $F$ and $y_m^+$ remains below $0.2$ at all times. For comparison, figure~\ref{fig:corr-F-ymp}(b) presents the correlation between $\ln(k_{ns}/E)$ and $y_m^+$, which is significantly stronger.
Following \citet{osawa2024causal}, $y_m^+$ is used to characterize the wall-normal position of the structures, as it is closely associated with the wall-attachment process. Alternative definitions, such as the energy-weighted centroid of the perturbation, yield qualitatively similar results. The near-symmetry of the probability density functions about $F=0$ further indicates that, at each wall-normal location, the probability of observing a structure whose causality is enhanced by the dynamics is approximately equal to that of observing one whose causality is suppressed.

\subsection{Key features for causally enhanced/suppressed events}

\begin{table}
\centering
\begin{tabular}{ll|ll}
Symbol & Description & Symbol & Description \\

$y_m^+$ & Minimum wall-normal location 
& $\Delta(u_i^2)$ & Velocity fluctuation energy \\

$y_g^+$ & Gravity center in wall-normal direction 
& $\Delta(uv)$ & Reynolds shear stress \\

$\Delta y^+$ & Wall-normal extent 
& $\Delta\!\left((uv)^2\right)$ & Squared Reynolds stress \\

$l_v^+$ & Volume-based length scale 
& $\Delta(\omega_i^2)$ & Enstrophy (no summation) \\

$l_s^+$ & Surface-based length scale 
& $\Delta\!\left(uv\,\partial_y U\right)$ & Production of turbulent kinetic energy \\

$\langle\partial_y U\rangle_c$ & Mean shear
& $\Delta\!\left(\nu \nabla^2 u_i^2\right)$ & Viscous diffusion (no summation) \\

$\langle\partial_y U\rangle_{fw}$ & Mean shear using only $y^+>5$
& $\Delta\!\left(\partial_y u\right)$ & Wall-normal velocity gradient \\

& 
& $\Delta\!\left(\partial_y u\right)_{fw}$ & Wall-normal gradient using only $y^+>5$ \\

& 
& $\Delta\!\left(\partial_i u_i\right)$ & Continuity-related term (no summation) \\

& 
& $\Delta(\omega_i)$ & Vorticity \\

& 
& $\Delta(S^2)$ & Strain-rate magnitude \\

& 
& $\Delta(u_i)$ & Velocity \\

& 
& $\Delta(P)$ & Initial production of perturbation energy \\

& 
& $\Delta(\varepsilon)$ & Initial dissipation of perturbation energy \\
\end{tabular}
\caption{Geometrical (left) and dynamical (right) features characterizing the structures. Detailed definitions are provided in the main text.}
\label{tab:features}
\end{table}

In this section, we identify the key features that determine whether a given event is causally enhanced or suppressed. To this end, we employ a support vector machine (SVM) for classification.

We first define a set of features characterizing each structure, as summarized in table \ref{tab:features}. These features are divided into two groups. The first one comprises geometrical features, including the wall-normal location and size of the structures.
{
Specifically, we consider two measures of location: the minimum wall-normal distance $y_m^+$ and the wall-normal centroid
\begin{equation}
    y_c^+ = \frac{1}{V_Q}\int_{Q} \, y^+ \, dV,
\end{equation}
together with three measures of size: the wall-normal extent $\Delta y^+=(y_M-y_m)^+$, the volume-based length scale 
\begin{equation}
    l_v=V_Q^{1/3},
    \label{eq:lv}
\end{equation}
and the surface-based length scale
\begin{equation}
    l_s=\left(\frac{V_Q}{y_M-y_m}\right)^{1/2}.
\end{equation}
The mean shear is also included in this group, as it is primarily determined by the wall-normal position. All geometrical features remain unchanged under the shifting operation.
}

The second group consists of dynamical features, which differ between the non-shifted and shifted structures. We characterise these differences by the normalised quantity
\begin{equation}
    \Delta \xi = \frac{\xi_{ns} - \xi_s}{\langle \xi_{\mathrm{rms}} \rangle_c},
\end{equation}
where $\xi_{ns}$ and $\xi_s$ denote the feature values for the non-shifted and shifted structures, respectively, and
\begin{equation}
    \langle \xi_{\mathrm{rms}} \rangle_c = \frac{1}{V_Q} \int_Q \xi_{\mathrm{rms}} \, \mathrm{d}V,
\end{equation}
is the volume-averaged root-mean-square value. This normalization quantifies the expected variation between structures with identical geometry but different streamwise and spanwise positions.
{For $P$ and $\varepsilon$, which are defined in \eqref{eq:prod} and \eqref{eq:diss}, the normalizing factor is set to wall units.
}

Following \citet{jimenez2020monte} and \citet{encinar2023identifying}, the ranking of observables is performed using a linear-kernel support vector machine (SVM) \citep{cristianini2000introduction}, implemented in the \texttt{scikit-learn} Python library \citep{pedregosa2011scikit}. The SVM determines an optimal separating hyperplane between two pre-labelled classes. In the present case, we seek to distinguish causally enhanced and causally suppressed events using a single observable, such that the SVM decision boundary reduces to a scalar threshold.
For each structure type (wall-attached Q2, wall-attached Q4, etc.), at least 240 samples are available for each class. One third of the samples are randomly selected as the training set, and the remaining two thirds are used for testing. 
{ The classification accuracy is defined as the fraction of correctly classified samples.
Because the numbers of the causally enhanced events and the causally suppressed events are similar, the classification is balanced, and $0.5$ corresponds to random guessing, whereas $1$ indicates perfect classification. 
{
To provide a simple interpretation of this score, consider two classes represented by identical normal distributions whose means are separated by $n$ standard deviations. Separations of one, two, three and four standard deviations correspond to approximate classification accuracies of 0.7, 0.84, 0.93 and 0.98, respectively.
}
Thus, the accuracy measures the capability of a given feature to discriminate between causally enhanced and causally suppressed events.
}
This random splitting is repeated three times, and the reported results correspond to the average accuracy. 

\begin{figure}
    \centering
    \includegraphics[width=\linewidth]{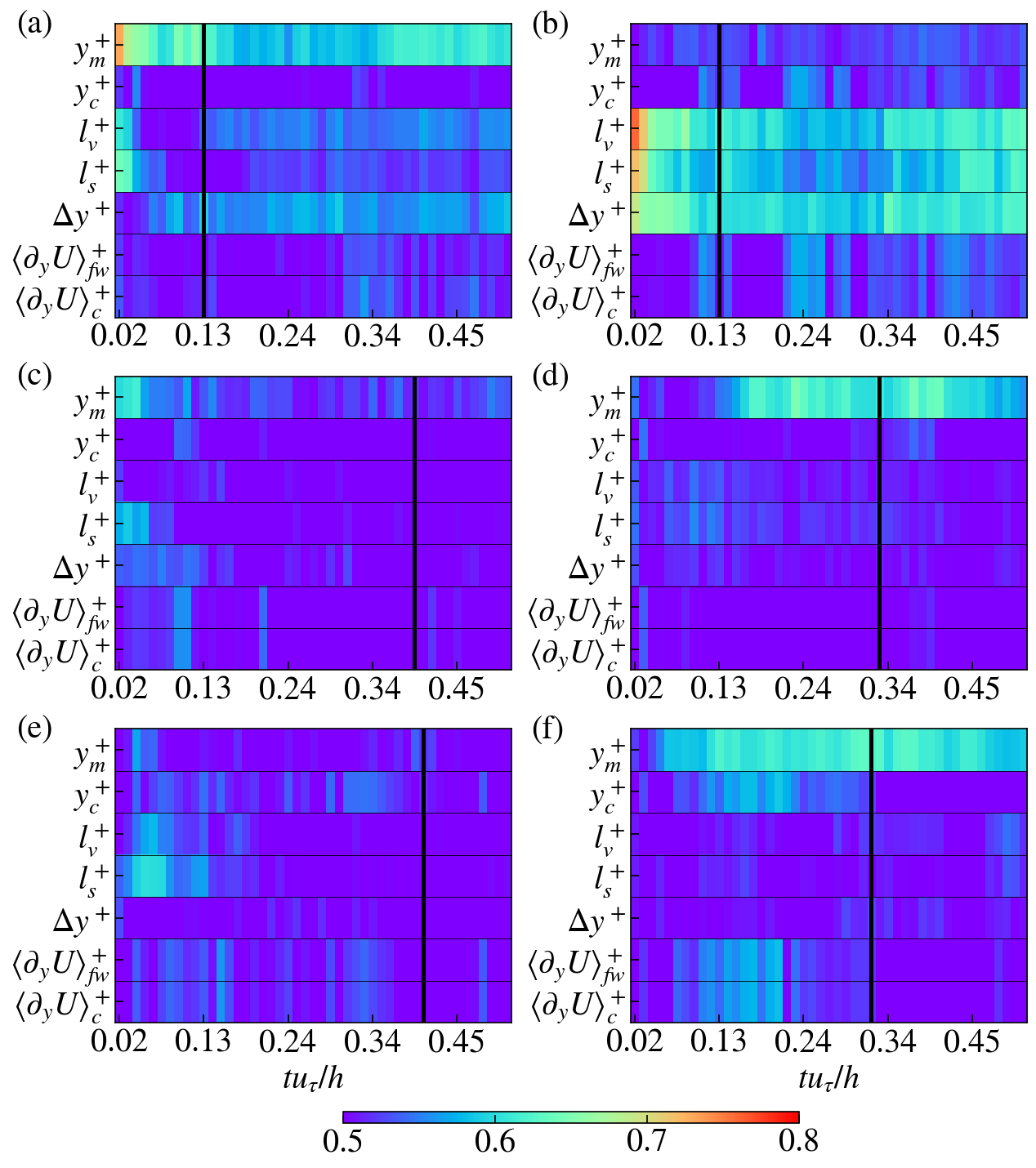}
    \caption{Classification accuracy of geometrical features for (a) wall-attached Q2, (b) wall-attached Q4, (c) wall-detached Q1, (d) wall-detached Q2, (e) wall-detached Q3, and (f) wall-detached Q4. The black solid lines are at $t_d$.}
    \label{fig:geo-feature}
\end{figure}

Figure \ref{fig:geo-feature} shows the classification performance of the geometrical features. Several observations can be made. First, geometrical features exhibit limited classifying capability, with mean accuracies below $0.6$.
{It is consistent with the results in figure \ref{fig:jpdf-F-ymp}, where $F$ is found to be independent of $y_m^+$.}
Second, for wall-attached Q2, wall-detached Q2, and wall-detached Q4, location-based features show classifying capability, whereas size-related features do not. In contrast, for wall-attached Q4, size-related features are more informative, while location plays a minor role.
The mean shear shows negligible classifying power in all cases.
These differences are likely related to the underlying dynamical features, which will be discussed below.
Third, the peak classification accuracy occurs at early times for wall-attached Q2 and Q4, whereas for wall-detached Q2 and Q4 it occurs at later times (around $0.3h/u_\tau$). This behavior is consistent with the distinguished time $t_d$ defined in figure \ref{fig:F-qs}.

\begin{figure}
    \centering
    \includegraphics[width=\linewidth]{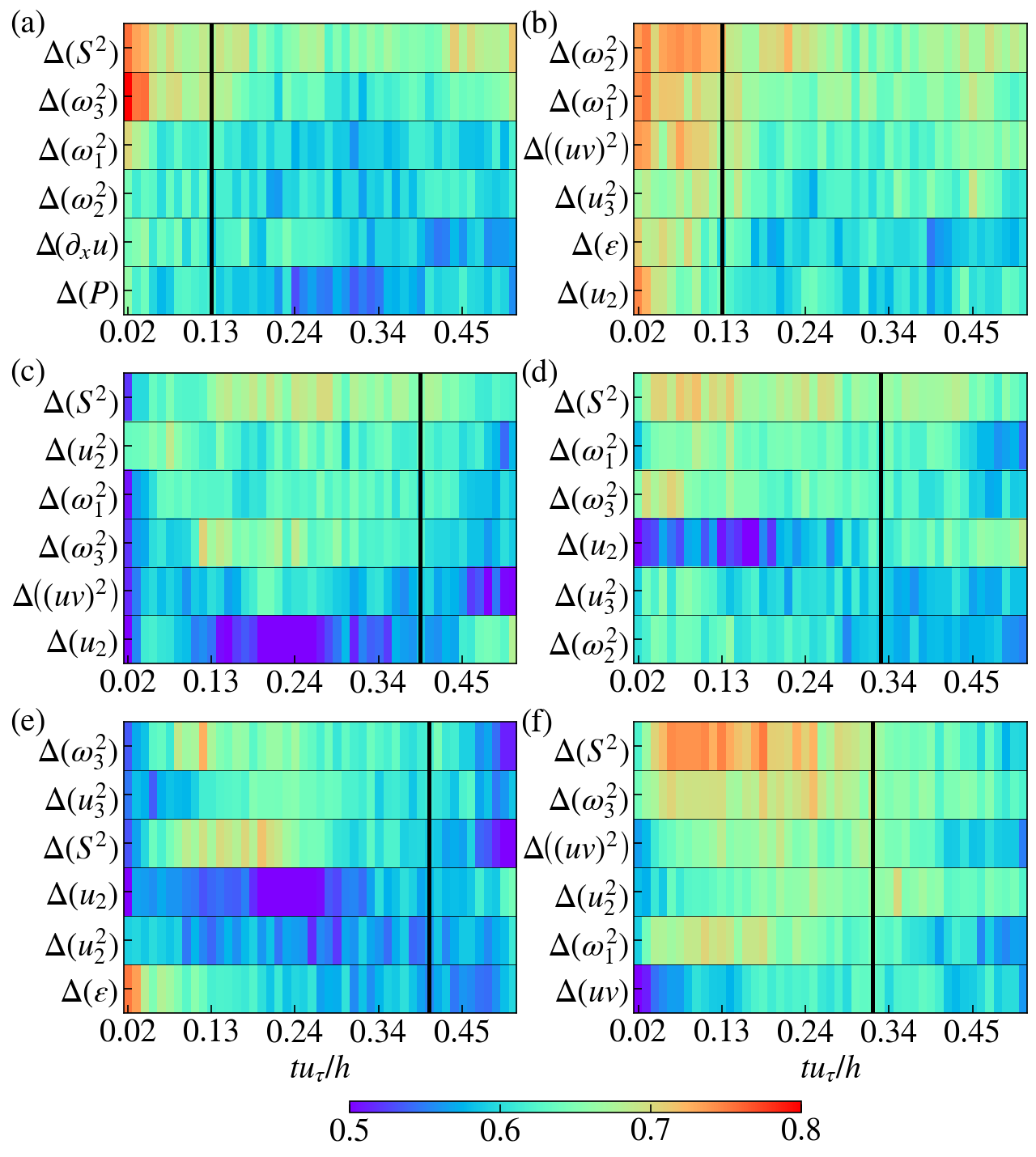}
    \caption{Classification accuracy of dynamical features for (a) wall-attached Q2, (b) wall-attached Q4, (c) wall-detached Q1, (d) wall-detached Q2, (e) wall-detached Q3, and (f) wall-detached Q4. The black solid lines are at $t_d$.}
    \label{fig:dyn-feature}
\end{figure}

Figure \ref{fig:dyn-feature} shows the classification performance of the dynamical features. For brevity, only the six most informative features are displayed, sorted by the accuracy at $t_d$.
Several observations can be made. First, the classification capability is significantly stronger than that of the geometrical features, with the most informative feature achieving a mean accuracy of approximately $0.75$.
Second, the dominant features depend on the structure type. For wall-attached Q2 events, the most informative features are the strain-rate magnitude and the spanwise vorticity. In contrast, for wall-attached Q4 events, the wall-normal and streamwise vorticity components provide the highest classification accuracy.
{ Strain and vorticity have also been identified as key features, second only to kinetic energy, in determining the causal significance of randomly distributed regions in homogeneous isotropic turbulence \citep{encinar2023identifying}.}
Third, the peak classification accuracy occurs at early times for wall-attached Q2 and Q4 events, whereas for wall-detached cases it occurs at later times, although the temporal variation is less pronounced. 

\begin{figure}
    \centering
    \includegraphics[width=\linewidth]{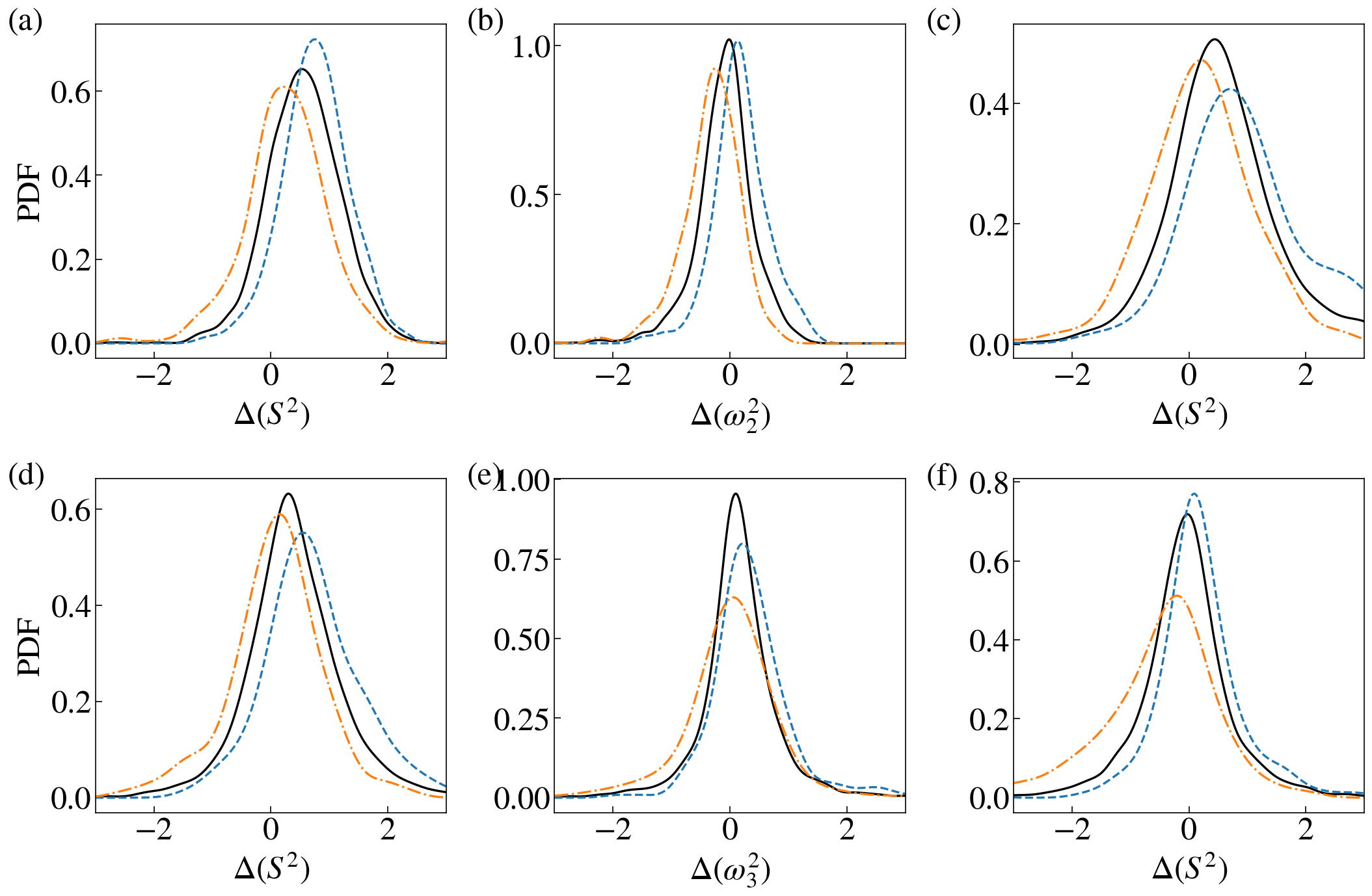}
    \caption{Probability density functions (PDFs) of the most informative feature for (a) wall-attached Q2, (b) wall-attached Q4, (c) wall-detached Q1, (d) wall-detached Q2, (e) wall-detached Q3, and (f) wall-detached Q4. The black solid lines denote the unconditional distributions, while the blue dashed and yellow dash-dotted lines correspond to the distributions conditioned on causally enhanced and suppressed events, respectively.}
    \label{fig:pdf-feature}
\end{figure}

Figure \ref{fig:pdf-feature} shows the PDFs of the most informative dynamical feature. As expected, the distribution of $\Delta(S^2)$ is skewed towards positive values, reflecting the elevated strain rate and dissipation associated with quadrant events. This skewness is weaker for wall-detached cases than for wall-attached ones. In particular, the distribution for wall-detached Q4 is approximately symmetric.
For wall-attached Q4, the PDF of $\Delta(\omega_2^2)$ exhibits a slight bias towards negative values, suggesting a weak correlation between wall-normal vorticity and Reynolds shear stress.
To further elucidate the distinction between causally enhanced and suppressed events, conditional PDFs are examined. As shown in figure \ref{fig:pdf-feature}, causally enhanced events are associated with larger values of $\Delta(S^2)$ in all cases except wall-attached Q4 and wall-detached Q3, whereas causally suppressed events preferentially correspond to smaller values. This indicates that strong strain is a key requirement for these quadrant events to be causally significant.
For wall-attached Q4, causally enhanced events are associated with positive $\Delta(\omega_2^2)$, while causally suppressed events tend to exhibit negative values. This suggests that an excess of wall-normal vorticity relative to the background turbulence is required for these structures to be causally significant.

\subsection{Flow structures around causally enhanced/suppressed events}
\label{sec:cond-u}
In the previous section, we showed that wall-attached Q2, wall-detached Q2 and wall-detached Q4 structures are more likely to be causally significant when they exhibit strong strain, whereas wall-attached Q4 structures are associated with strong wall-normal vorticity. In this section, we further investigate these relationships by analyzing conditionally averaged flow fields.
The velocity field is first shifted such that each structure is centered at its centroid in the streamwise and spanwise directions, defined as
\begin{equation}
    x_c = \frac{1}{V_Q}\int_Q x \,\mathrm{d}V, \quad 
    z_c = \frac{1}{V_Q}\int_Q z \,\mathrm{d}V.
    \label{eq:xc-zc}
\end{equation}
The velocity fields are then ensemble-averaged.
{

To avoid spanwise symmetry in the conditional averages \citep{stretch1990automated, lozano2012three, dong2017coherent, osawa2024causal}, the spanwise orientation of each structure is first standardized. For each structure, we compute the mean spanwise velocity in a slightly enlarged domain surrounding the structure. This domain is defined as the set of points satisfying $|x-Q|^+<30$ and $|z-Q|^+<30$, where $|x-Q|$ and $|z-Q|$ denote the streamwise and spanwise distances from the point to the structure domain $Q$, respectively. 
{
The cutoff distance of 30 wall units is equal to the smallest size of the structures. 
Note that the structures with  $l_v^+<30$ have been removed, where $l_v^+$ is the volume-based size defined in  \eqref{eq:lv}. 
}
If the mean spanwise velocity in this enlarged domain is negative, a mirror transformation is applied in the spanwise direction, namely $z\to -z$ and $w\to -w$.
}
For wall-detached Q2 and Q4 events, the conditionally averaged fields depend on the wall-normal location of the structures, and we focus on events with $50<y_m^+<150$. 
{
Table~\ref{tab:num-vol-cond} lists the numbers and volume fractions of the structures used for the conditional averages. For wall-attached Q2 and Q4 events, the sample sizes are sufficiently large for the conditional averages to be well converged. For wall-detached Q2 and Q4 events, the sample sizes are smaller, and the corresponding conditionally averaged fields are therefore noisier.

{
\begin{table}
    \centering
    \begin{tabular}{ccccccccccccc}
         Structure type & $N_{\mathrm{e}}$ & $N_{\mathrm{s}}$ & $V_{\mathrm{e}}$ & $V_{\mathrm{s}}$ & $V_{85,\mathrm{e}}$ & $V_{85,\mathrm{s}}$ & $l_{85,e}^+$ & $l_{85,s}^+$  & $R_e$  & $R_s$\\
         Attached Q2 & 263 & 263 & 0.034 & 0.067  & 0.29 &  0.13 & 33.7  & 32.2 &  0.34 &  0.67 \\
         Attached Q4 & 235 & 235 & 0.115 & 0.018  & 0.07 &  0.40 & 33.7  & 32.5 &  1.14 &  0.18 \\
         Detached Q2 & 68  & 142 & 0.008 & 0.016  & 0.24 &  0.24 & 32.2  & 32.5 &  0.38 &  0.39 \\
         Detached Q4 & 113 & 53  & 0.014 & 0.004  & 0.34 &  0.42 & 39.7  & 36.1 &  0.48 &  0.30
    \end{tabular}
    \caption{Number and volume fractions of the structures used for the conditional averages. Here, $N$ denotes the number of structures and $V$ denotes their volume fraction relative to the total volume of all structures of the same type. The quantity $V_{85}$ denotes the fraction of the total structure volume contained within the region enclosing 85\% of the structures (\eqref{eq:V85}). $l_{85}=(\int_{\rho>0.15} \rho dV)^{1/3}$ is the size of the region enclosing 85\% of the structures.  $R$ is the volume ratio between the mean volume of the conditional samples and that of the all structures of the same type. The subscripts ${e}$ and ${s}$ denote causally enhanced and causally suppressed structures, respectively.}
    \label{tab:num-vol-cond}
\end{table}

\begin{figure}
    \centering
    \includegraphics[width=0.66\linewidth]{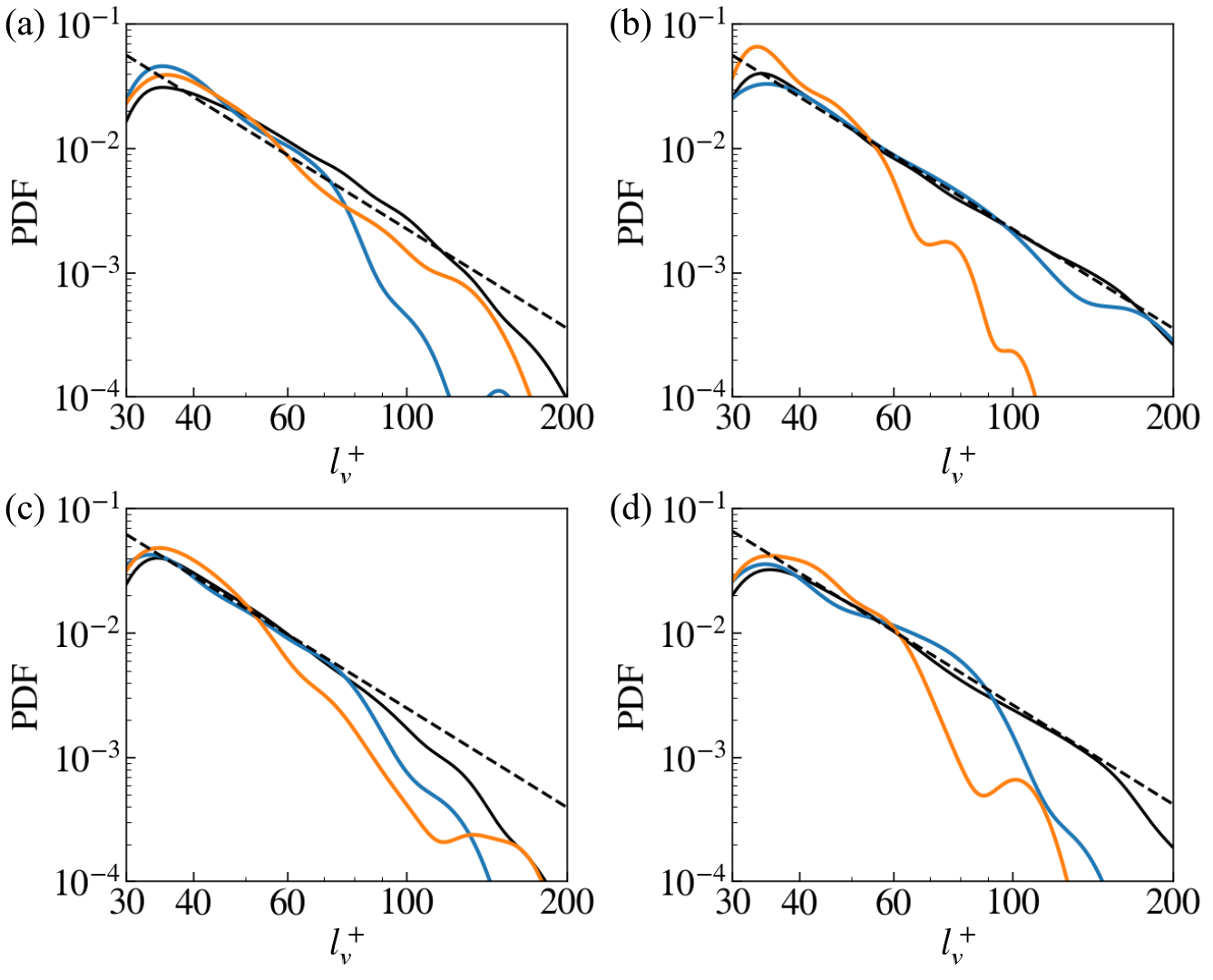}
    \caption{PDF of the volume-based length scale $l_v^+$ for (a) wall-attached Q2, (b) wall-attached Q4, (c) wall-detached Q2 and (d) wall-detached Q4 structures. The black solid lines correspond to all structures of the same type. The blue and orange solid lines correspond to causally enhanced and causally suppressed structures, respectively. The black dashed lines stand for the scaling $l_v^{-2.67}$. For the wall-detached cases, only structures with $50<y_m^+<150$ are included.}
\label{fig:pdf-lv}
\end{figure}

\begin{figure}
    \centering
    \includegraphics[width=\linewidth]{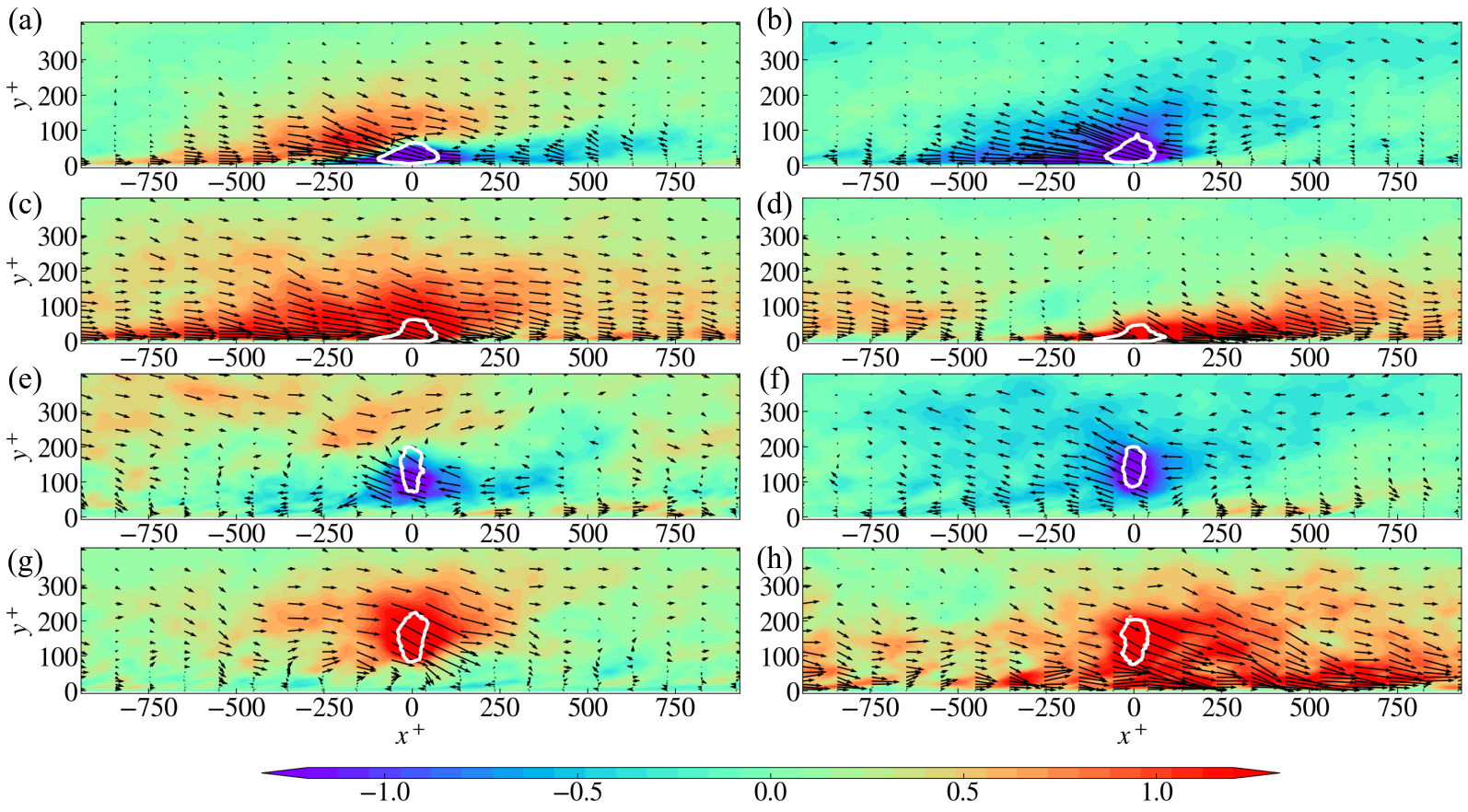}
    \caption{$xy$-slice of the conditionally averaged velocity field for (a,b) wall-attached Q2, (c,d) wall-attached Q4, (e,f) wall-detached Q2, and (g,h) wall-detached Q4. Panels (a,c,e,g) correspond to causally enhanced structures, and (b,d,f, h) to causally suppressed structures. Contours show $u$, and vectors indicate the velocity field. The white contours contain $85\%$ of the structures. }
    \label{fig:cond-v-xy}
\end{figure}

\begin{figure}
    \centering
    \includegraphics[width=\linewidth]{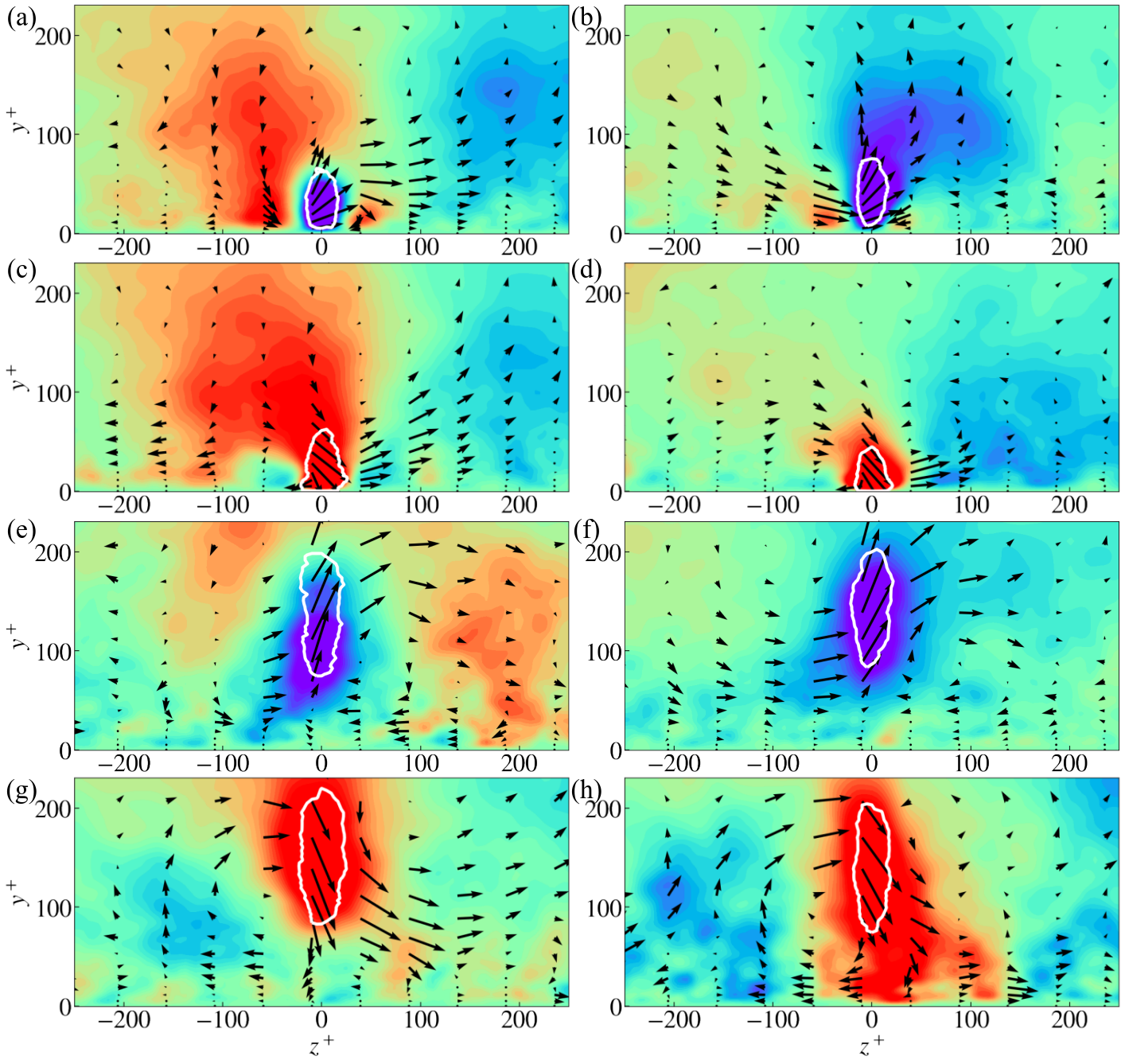}
    \caption{$yz$-slice of the conditionally averaged velocity field. Panels and notation as in figure \ref{fig:cond-v-xy}.}
    \label{fig:cond-v-yz}
\end{figure}

\begin{figure}
    \centering
    \includegraphics[width=\linewidth]{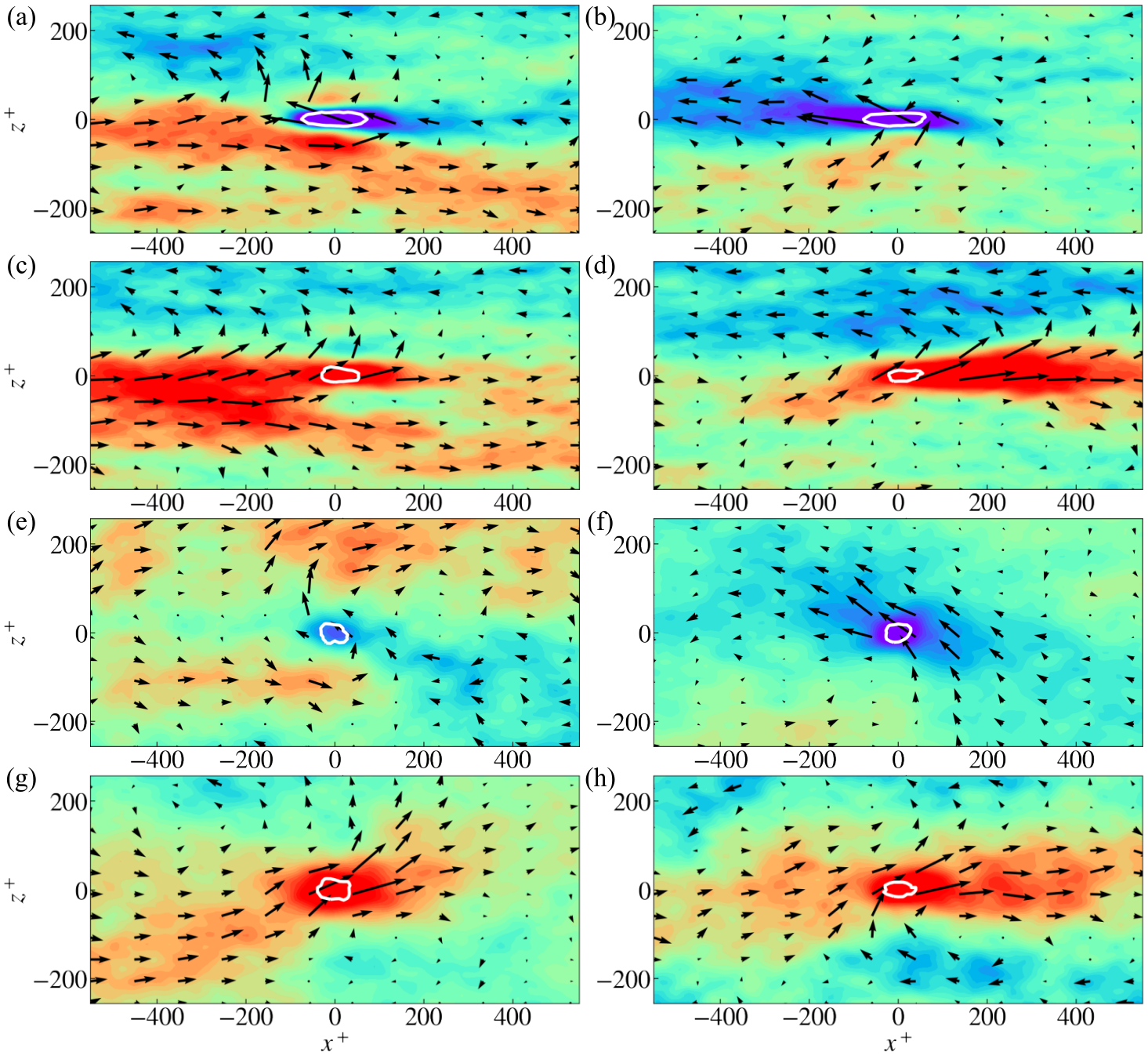}
    \caption{$xz$-slice of the conditionally averaged velocity field. Panels and notation as in figure \ref{fig:cond-v-xy}. (a-d) The slice is taken at $y^+=30$. (e-f) The slice is taken at $y^+=150$.}
    \label{fig:cond-v-xz}
\end{figure}


Because the structures have complex geometries, especially when their sizes are large, defining a representative boundary in the conditionally averaged fields is not straightforward. To do so, we introduce the occupancy density
\begin{equation}
\rho(\boldsymbol{x})
=
\frac{1}{N_t}
\sum_{i=1}^{N_t}
\mathbf{1}_{Q_i^s}(\boldsymbol{x}),
\label{eq}
\end{equation}
where $N_t$ is the total number of structures in the conditional sample, $Q_i^s$ denotes the domain occupied by the $i$th structure after translation to center at $(x_c,z_c)$, and $\mathbf{1}_{Q_i^s}$ is the characteristic function of $Q_i^s$. 
$\rho=0$ is the envelope of the translated structures.
However, this envelope is strongly affected by small-scale branches and rare protrusions of individual structures, and is therefore not a robust representation of the boundary between the conditioned structures and their environment.
As shown in Appendix \ref{app:cond},  $\rho=0.15$ provides a suitable representation of the boundary between conditioned structures and their environment.
The region enclosed by $\rho=0.15$ contains approximately $85\%$ of the total structures, and the corresponding volume fraction of the region
\begin{equation}
V_{85}
=
\frac{
\displaystyle
\int_{\rho>0.15}
\rho(\boldsymbol{x})\mathrm{d}V
}{
\displaystyle
\int
\rho(\boldsymbol{x})\mathrm{d}V
},
\label{eq:V85}
\end{equation}
 are reported in Table~\ref{tab:num-vol-cond}. 
Moreover, the characteristic size of the 85\% regions $l_{85}^+=(\int_{\rho>0.15} \rho dV)^{1/3} u_\tau/\nu=30\sim 40 $.

It is noteworthy that the mean volume of causally enhanced wall-attached Q4 structures is comparable to that of the full population of wall-attached Q4 structures. In the other cases, however, the mean volume of the structures in the conditional samples, including both causally enhanced and causally suppressed structures, is smaller than that of the corresponding full population.
This size bias is further illustrated in figure~\ref{fig:pdf-lv}, which shows the PDF of the volume-based length scale $l_v^+$ defined in \eqref{eq:lv}. For wall-attached Q2 structures, the PDF of the causally suppressed events is only slightly shifted towards smaller values of $l_v^+$ relative to the full population, whereas the PDF of the causally enhanced sample differs more strongly. This indicates that, for wall-attached Q2 events, causal enhancement occurs preferentially among smaller structures.
For wall-attached Q4 structures, the PDF of the causally enhanced sample is close to that of the full population, indicating that the distribution of $l_v^+$ is relatively insensitive to whether the structures are causally enhanced. This does not contradict the feature-analysis results, because the distribution of $l_v^+$ does depend on whether the structures are causally suppressed.
For wall-detached Q2 and Q4 structures, both the causally enhanced and causally suppressed PDFs are shifted towards smaller values of $l_v^+$. However, the causally enhanced PDFs remain closer to the full-population PDFs than the causally suppressed PDFs.
Finally, \citet{lozano2012three} showed that the PDF of the wall-normal height of wall-attached structures follows the scaling $\Delta y^{-2}$, while the relation between volume and height is approximately $V_Q\sim \Delta y^{2.25}$ for wall-attached Q2 and Q4 structures as shown in figure \ref{fig:urms-vol-ymin}(c). 
{These scalings imply that the PDF of $l_v$ follows the scaling
\begin{equation}
    \mathrm{PDF}\sim \Delta y^{-2} \sim V_Q^{-2/2.25} \sim (l_v^+)^{-3\times 2/2.25} =(l_v^+)^{-2.67}.
\end{equation}
As shown in figure~\ref{fig:pdf-lv}, the PDF of $l_v^+$ follows this scaling reasonably well, even for wall-detached Q2 and Q4 restricted to $50<y_m^+<150$.

}
}
{
Figures~\ref{fig:cond-v-xy}, \ref{fig:cond-v-yz} and \ref{fig:cond-v-xz} show, respectively, the $xy$-, $yz$- and $xz$-slices of the conditionally averaged velocity fields for wall-attached and wall-detached Q2 and Q4 events.
As shown in figures~\ref{fig:cond-v-xy}(a,b), the flow fields conditioned on causally enhanced and causally suppressed wall-attached Q2 events are markedly different. For causally enhanced Q2 events, an upstream Q4-like motion impinges on the Q2 region and is deflected around it. 
A strong { strain region} forms between the two motions. The inclination angle of this { strain region} is approximately $14^\circ$. These high-strain regions are then advected downstream, leading to Q2 events with enhanced strain and spanwise vorticity, consistent with figures~\ref{fig:dyn-feature} and \ref{fig:pdf-feature}. By contrast, for causally suppressed Q2 events, the upstream motion is also Q2-like, and no comparable { strain region} is observed.
Figures~\ref{fig:cond-v-xy}(c,d) show that no strong { strain region} is present for either causally enhanced or causally suppressed wall-attached Q4 events. This is consistent with the observation that $\Delta(S^2)$ is not a dominant feature for Q4 structures in figure~\ref{fig:dyn-feature}. For causally enhanced Q4 events, the upstream flow is Q4-like, whereas for causally suppressed Q4 events the upstream velocity fluctuation is weak.
For wall-detached Q2 events, figures~\ref{fig:cond-v-xy}(e,f) show that causally enhanced events are accompanied by shear layers both above and below the conditioned structure, whereas causally suppressed events exhibit only a weaker shear layer below it.
For wall-detached Q4 events, figures~\ref{fig:cond-v-xy}(g,h) show that the main difference lies in the velocity field below the conditioned structure. In the causally enhanced case, a weak low-speed region is present underneath the Q4 event, whereas in the causally suppressed case the velocity below the event remains Q4-like.

The low-speed object located beneath the high-speed overhang in figure~\ref{fig:cond-v-yz}(a) resembles the low-speed ramp structures previously observed in streamwise sections of wall turbulence \citep[e.g.][]{meinhart1995existence, lozano2012three}. A similar, although weaker, ramp-like structure is also observed for causally enhanced wall-attached Q4 events in figure~\ref{fig:cond-v-yz}(c). For wall-detached events, several detached vortical structures are observed in the conditionally averaged fields.

Finally, figure~\ref{fig:cond-v-xz} shows the organization of the conditioned events relative to streaky motions \citep{tomkins2003spanwise, lozano2012three}. Causally enhanced Q2 events tend to occur near the upstream end of low-speed streaks, whereas causally suppressed Q2 events are more commonly located toward their downstream end. Conversely, causally enhanced Q4 events tend to lie near the downstream end of high-speed streaks, while causally suppressed Q4 events are more often found near their upstream end. These trends further support the view that the causal significance of quadrant events is controlled not only by the events themselves, but also by their position within the surrounding streak organization.

It is useful to compare figures~\ref{fig:cond-v-xy}, \ref{fig:cond-v-yz} and \ref{fig:cond-v-xz} with figures 13--15 of \citet{osawa2024causal}. In their randomly sampled interventional experiments, causally significant cells were found to move downstream towards an interface where high-speed fluid overtakes low-speed fluid. By contrast, causally insignificant cells were associated with an interface where low-speed fluid is left behind by faster fluid located downstream. 
The present results show that a similar overtaking process is important for the causal significance of quadrant events, particularly for wall-attached and wall-detached Q2 structures. 
}

\section{Conclusions}
\label{sec:conclusions}

The importance of quadrant events has been examined from a causal perspective. We employed a pairwise interventional framework in which, in the treated cases, the velocity field was modified within regions occupied by quadrant events, while in the control cases the modification was applied to regions with identical shape and wall-normal location but shifted in the wall-parallel directions. 
{
The causal significance of the non-shifted and shifted structures was quantified through the subsequent evolution of the perturbation energy. Their ratio was then used to measure the causal advantage of quadrant events relative to the background turbulence.
}

We find that for all four types of quadrant events, namely Q1, Q2 (ejection), Q3, and Q4 (sweep), their coherent dynamics does not always enhance their causal significance. The number of causally suppressed events is comparable to that of causally enhanced events.

This, however, does not imply that the causal significance is independent of the coherent dynamics. On the contrary, the SVM-based feature analysis reveals clear dynamical signatures. For wall-attached Q2 and wall-detached structures, the strain rate and spanwise vorticity are the key features distinguishing causally enhanced from suppressed events. Causally enhanced events are associated with higher levels of strain and spanwise vorticity.
The conditionally averaged velocity fields show that this elevated strain and spanwise vorticity originate from an intense upstream strain region, formed by the interaction with an upstream Q4-like structure impinging on the Q2 event.
For wall-attached Q4 structures, the dominant features are the wall-normal and streamwise vorticity components: {Causally enhanced Q4 events preferentially exhibit larger wall-normal vorticity, whereas causally suppressed events are associated with lower values.}

{
These results should be compared with previous interventional studies of causally significant flow regions. In wall-bounded turbulence, \citet{osawa2024causal} found that wall-normal velocity is one of the most important features distinguishing significant randomly selected regions. In homogeneous isotropic turbulence, \citet{encinar2023identifying} found that kinetic energy is the leading feature, followed by quantities associated with $\nabla\boldsymbol{u}$, such as $\boldsymbol{\omega}^2$ and $S^2$. In the present analysis, the velocity-based features are partly constrained by construction, because the structures are selected as extreme events in $u$ and $v$. This may explain why features based on $\nabla\boldsymbol{u}$, such as strain and vorticity, become the leading discriminators of causal significance.
This interpretation suggests that velocity-based structures, such as quadrant events, and gradient-based structures, such as intense strain regions or vortex clusters, should not be regarded as independent objects. Instead, they adjust to one another and form compound flow configurations. The present results show that some of these configurations, such as those in our figure~\ref{fig:cond-v-xy}(a,c,e,g), are more causally significant than others, such as those in our figure~\ref{fig:cond-v-xy}(b,d,f,h).
}

{
Several limitations should be noted. In the present study, quadrant events were identified from 40 velocity fields separated by approximately 1.5 eddy-turnover times. The lifetime of Q2 and Q4 events is of the order of the local eddy-turnover time, $T\sim (y_M-y_m)/u_\tau$, which is shorter than the time interval between our consecutive snapshots \citep{lozano2014time}. The quadrant events considered here can therefore be treated as statistically independent individuals whose lifetimes do not overlap. However, the present analysis does not distinguish between different stages in the life cycle of the events. Some structures may be newly formed, whereas others may be close to decay. This limitation makes it difficult to determine whether differences in causal significance arise from the life-cycle stage of the structures at the time of perturbation, or from more intrinsic differences in their dynamics when their complete life cycle is considered.

In addition, the present study classifies quadrant events only as wall-attached or wall-detached. A finer classification may reveal further distinctions. For example, \citet{dong2017coherent}, among others, showed that tall wall-attached events differ from shorter ones, while most of our structures may be classified as short ($y^+_M<100$). However, a finer classification would require substantially more data to ensure statistical convergence, and there is no unique criterion for determining the appropriate level of refinement. The conclusions drawn here should therefore be interpreted within the classification framework adopted in this study.
}

Our results address the open question of whether all coherent structures are more causally significant than the background turbulence. The answer is clearly negative. For quadrant events, only those associated with intense strain or vorticity exhibit enhanced causal significance relative to the background turbulence.
Moreover, the surrounding flow environment plays a crucial role. This highlights that coherent structures should not be analyzed in isolation, but rather in the context of their dynamical interactions with the surrounding flow.
The present study represents an initial step towards understanding the relationship between coherence and causality in turbulence. Future work will extend this framework to other types of coherent structures and further elucidate the dynamical mechanisms underlying their causal influence.

\section*{Funding}
This work was supported by European Research Council under the Caust Grant ERC-AdG-101018287.

\section*{Declaration of interests}
The authors report no conflict of interest.

\section*{Data Availability Statement}
The data that support the findings of this study are available from the corresponding author upon reasonable request.

\appendix
\section{The influence of the pressure step}
\label{app:press}
\begin{figure}
    \centering
    \includegraphics[width=0.66\linewidth]{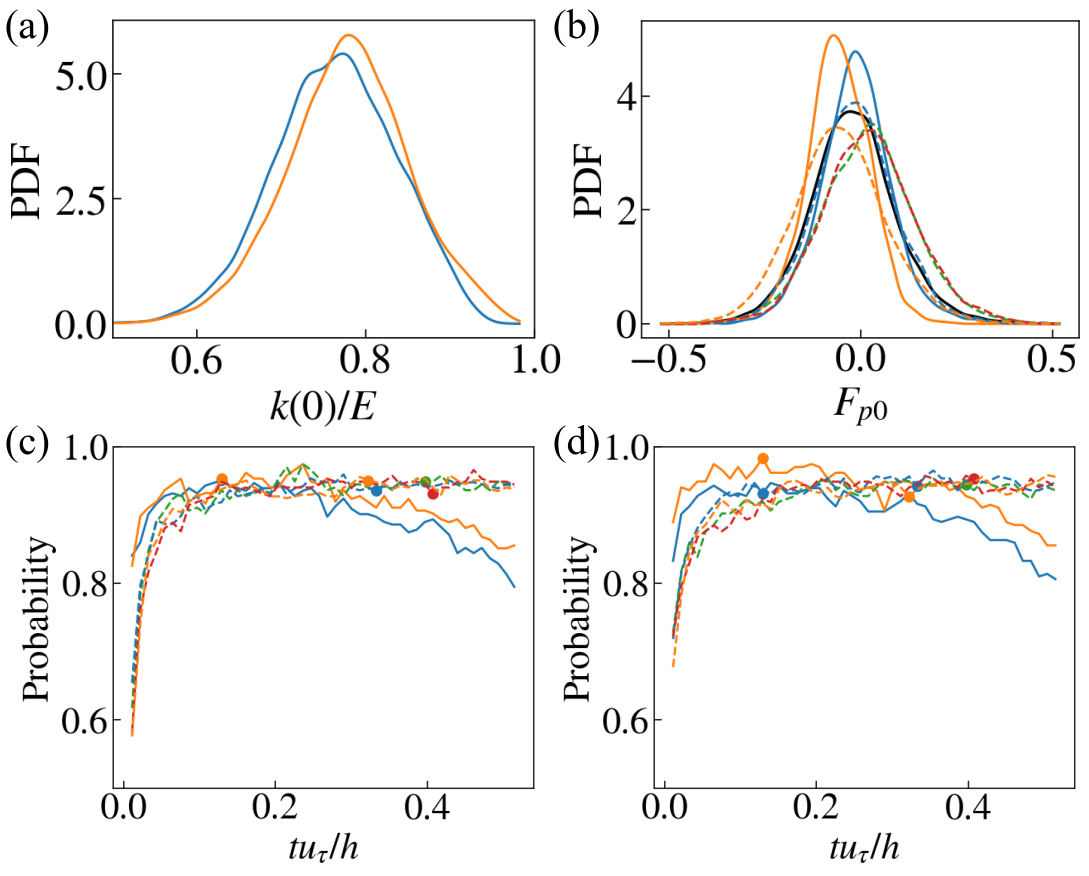}
    \caption{Effect of the pressure-projection step. 
    (a) Probability density function (PDF) of the perturbation energy at $t=0$, normalized by the prescribed perturbation energy $E$. Blue and yellow lines denote non-shifted and shifted structures, respectively. 
    (b) PDF of $F_{p0}$, defined in \eqref{eq:Fp0}. The black solid line denotes the unconditional PDF. The other lines denote conditional PDFs: blue solid line, wall-attached Q2; yellow solid line, wall-attached Q4; red dashed line, wall-detached Q1; blue dashed line, wall-detached Q2; green dashed line, wall-detached Q3; yellow dashed line, wall-detached Q4. 
    (c,d) Overlap probability defined in \eqref{eq:prob-pres-effect}. Panel (c) shows causally enhanced events, and panel (d) shows causally suppressed events. The line styles are the same as in (b).}
    \label{fig:press-effect}
\end{figure}
{
We examine the effect of the pressure-projection step. Although the perturbation is normalized to have prescribed energy $E$ before projection, the projection modifies the perturbation field and therefore changes its energy. Figure~\ref{fig:press-effect}(a) shows the PDF of the ratio $k(0)/E$, where $k(0)$ is the perturbation energy after pressure projection. The ratio lies mostly in the range $0.5<k(0)/E<1$, indicating that the pressure projection reduces the perturbation energy. The distributions for shifted and non-shifted structures are very similar, suggesting that the projection does not introduce a systematic bias between the two groups.

Because $k(0)$ differs from the prescribed value $E$, an alternative enhancement factor can be defined by normalizing the perturbation growth by its post-projection initial value:
\begin{equation}
    F_p(t)
    =
    \ln\left[
    \frac{k_{\mathrm{ns}}(t)/k_{\mathrm{ns}}(0)}
         {k_{\mathrm{s}}(t)/k_{\mathrm{s}}(0)}
    \right]
    =
    F(t)+F_{p0},
    \label{eq:Fp}
\end{equation}
where
\begin{equation}
    F_{p0}
    =
    \ln
    \frac{k_{\mathrm{s}}(0)}
         {k_{\mathrm{ns}}(0)} .
    \label{eq:Fp0}
\end{equation}
Figure~\ref{fig:press-effect}(b) shows that $F_{p0}$ is concentrated near zero and mostly lies within the range $-0.6<F_{p0}<0.6$. This indicates that the pressure-projection step has only a limited effect on the relative classification of shifted and non-shifted pairs.

To quantify the robustness of the classification, we compare the sets of events identified using $F$ and $F_p$. For example, for causally enhanced wall-attached Q2 events, we define the overlap probability as
\begin{equation}
    P_{\mathrm{ov}}^{+}(t)
    =
    \frac{
    P\left(
    \mathcal{T}_{Q2,att}^{F}(t)
    \cap
    \mathcal{T}_{Q2, att}^{F_p}(t)
    \right)
    }{
    P\left(
    \mathcal{T}_{Q2, att}^{F}(t)
    \right)
    },
    \label{eq:prob-pres-effect}
\end{equation}
where $\mathcal{T}_{Q2, att}^{F}(t)$ and $\mathcal{T}_{Q2, att}^{F_p}(t)$ denote the top $10\%$ of wall-attached Q2 events ranked by $F(t)$ and $F_p(t)$, respectively. An analogous definition is used for causally suppressed events, replacing the top $10\%$ by the bottom $10\%$.
Figures~\ref{fig:press-effect}(c,d) show the resulting overlap probabilities. Except at very early times, the overlap probability is larger than approximately $0.8$, confirming that the classification of causally enhanced and suppressed events is largely insensitive to whether the initial perturbation energy is normalized before or after the pressure projection. The overlap probability decreases more rapidly for wall-attached events than for wall-detached events, because the enhancement factor of wall-attached events approaches zero after $t=t_d\approx0.1h/u_{\tau}$.
}

\section{Conditional averaged velocity field within the structures}
\label{app:cond}
\begin{figure}
    \centering
    \includegraphics[width=\linewidth]{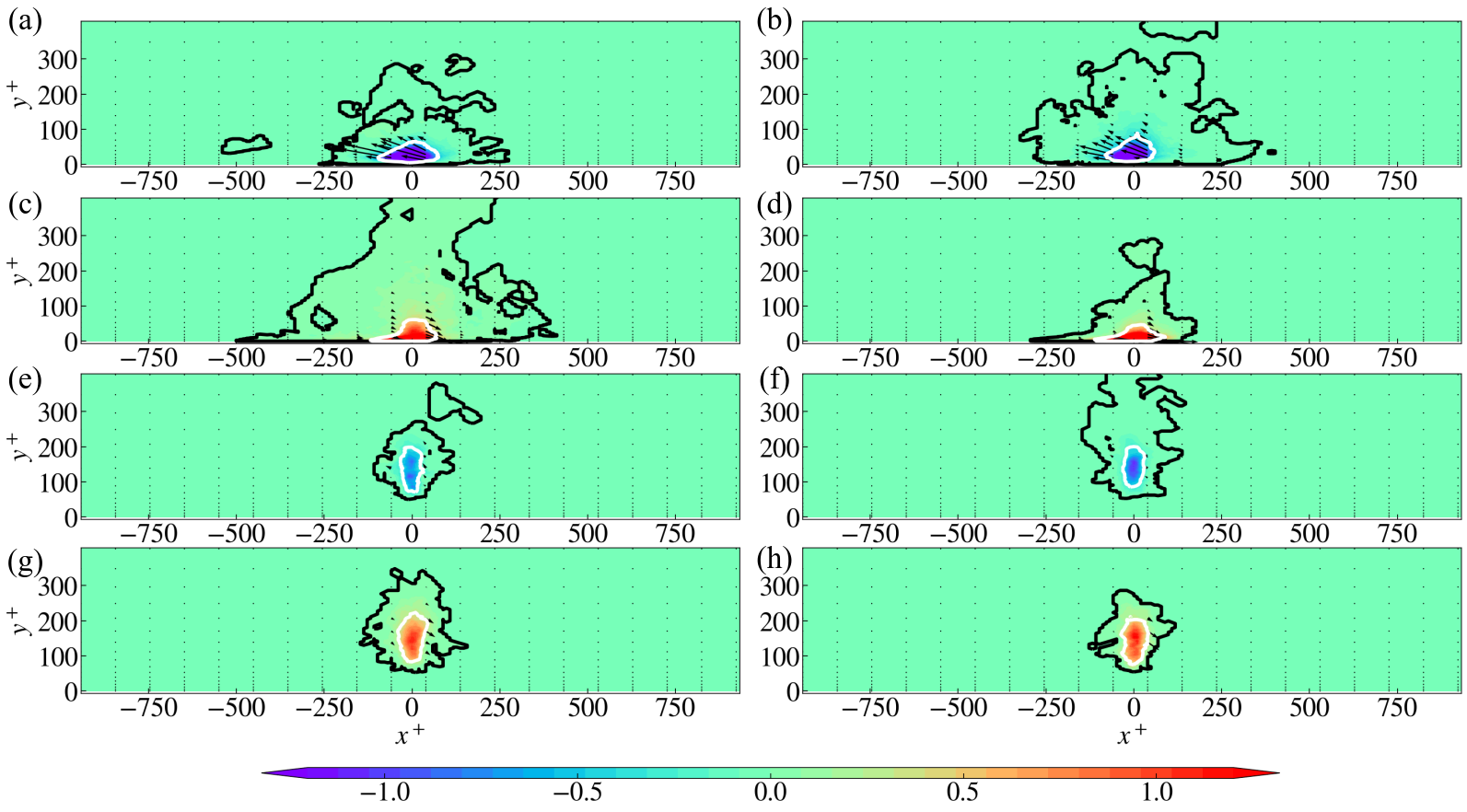}
    \caption{$xy$-slice of the conditionally averaged velocity field within the structures.  The black contours are the envelope of the structures, and the white contours contain 85\% of the structures. The other notations are the same as in figure \ref{fig:cond-v-xy}. }
    \label{fig:cond-v-xy-str}
\end{figure}

\begin{figure}
    \centering
    \includegraphics[width=\linewidth]{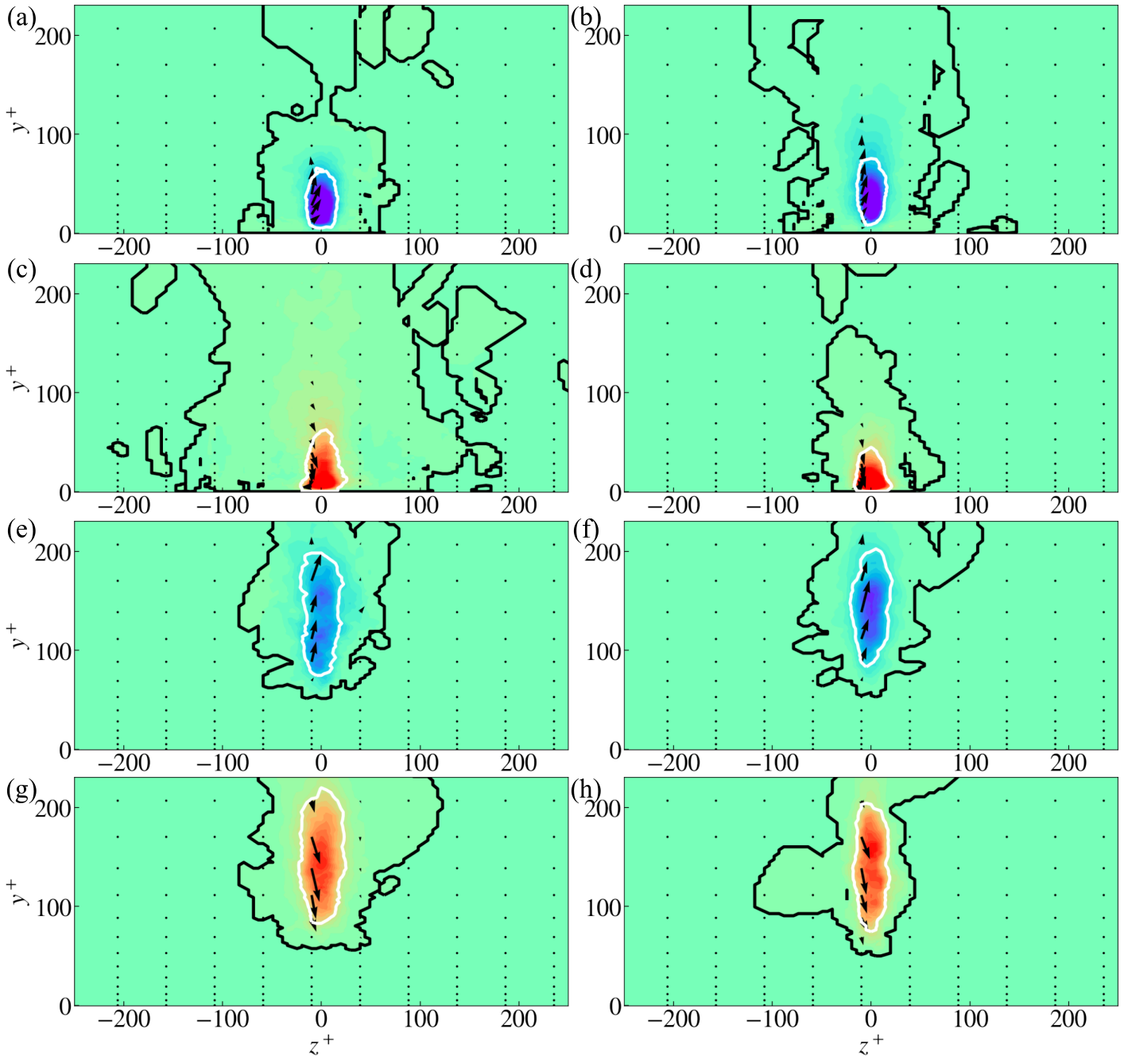}
    \caption{$yz$-slice of the conditionally averaged velocity field within the structures.  Panels and notation as in figure \ref{fig:cond-v-xy-str}.}
    \label{fig:cond-v-yz-str}
\end{figure}
\begin{figure}
    \centering
    \includegraphics[width=\linewidth]{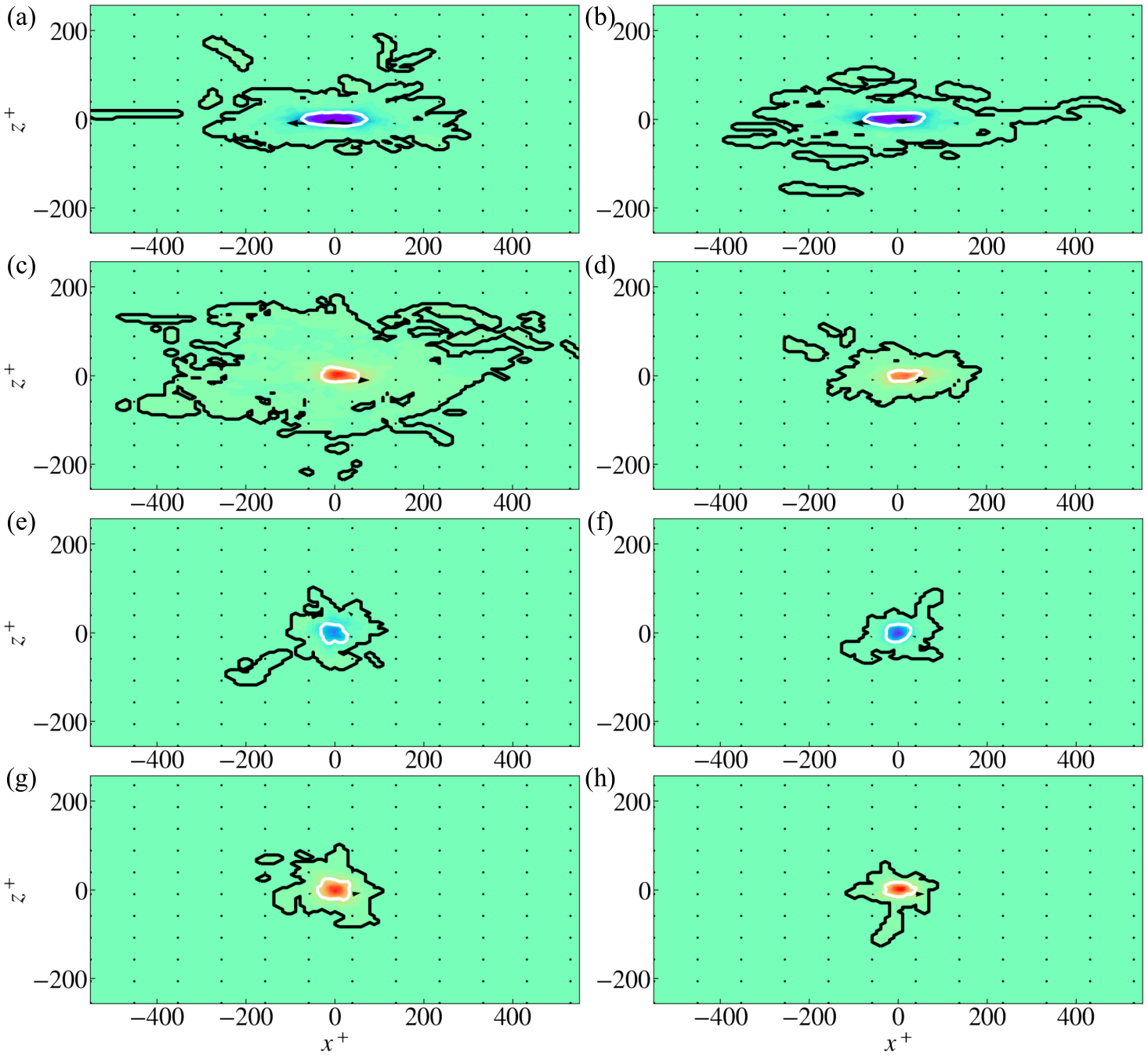}
    \caption{$xz$-slice of the conditionally averaged velocity field within the structures. Panels and notation as in figure \ref{fig:cond-v-xy-str}. }
    \label{fig:cond-v-xz-str}
\end{figure}

{

As described in \S~\ref{sec:cond-u}, the conditionally averaged velocity field is obtained by translating each instantaneous field so that the centroid of the corresponding structure is located at the origin in the streamwise and spanwise directions. Specifically,
\begin{equation}
\boldsymbol{u}_{ca}(\boldsymbol{x})
=
\frac{1}{N_t}
\sum_{i=1}^{N_t}
\boldsymbol{u}_i^s(\boldsymbol{x}),
\label{eq:uca}
\end{equation}
where the subscript $ca$ means conditionally averaged,  $N_t$ is the total number of structures in the conditional sample, and
$
\boldsymbol{u}_i^s(x,y,z)
=
\boldsymbol{u}(x+x_{c,i},y,z+z_{c,i}) .
$
Here, $x_{c,i}$ and $z_{c,i}$ are the streamwise and spanwise coordinates of the centroid of the $i$th structure before translation, as defined in \eqref{eq:xc-zc}.

The conditionally averaged field can be decomposed into contributions from the structures themselves and from their surrounding environment. Let $Q_i^s$ denote the translated domain occupied by the $i$th structure, and let $\mathbf{1}_{Q_i^s}$ be its characteristic function. We define
\begin{equation}
\boldsymbol{u}_{ca}
=
\frac{1}{N_t}
\sum_{i=1}^{N_t}
\boldsymbol{u}_i^s
\mathbf{1}_{Q_i^s}
+
\frac{1}{N_t}
\sum_{i=1}^{N_t}
\boldsymbol{u}_i^s
\left(
1-\mathbf{1}_{Q_i^s}
\right)
=
\boldsymbol{u}_{str}
+
\boldsymbol{u}_{env}.
\label{eq:uca-decomposition}
\end{equation}
Here, $\boldsymbol{u}_{str}$ represents the contribution from points inside the  structures, whereas $\boldsymbol{u}_{env}$ represents the contribution from the surrounding flow.

Figures~\ref{fig:cond-v-xy-str}, \ref{fig:cond-v-yz-str} and \ref{fig:cond-v-xz-str} show the corresponding fields $\boldsymbol{u}_{str}$ for wall-attached and wall-detached Q2 and Q4 events. The non-zero velocity regions are mostly contained within the $85\%$ contours, indicating that these contours provide a suitable representation of the common boundary between the structures and their environment. Outside the $85\%$ contours, $\boldsymbol{u}_{str}$ is close to zero, which implies that the conditionally averaged velocity field in those regions is dominated by the environmental contribution $\boldsymbol{u}_{env}$. Therefore, the coherent motions observed in figures~\ref{fig:cond-v-xy}, \ref{fig:cond-v-yz} and \ref{fig:cond-v-xz} primarily represent the surrounding environment rather than the conditioned structures themselves. This supports the conclusion that the causal enhancement of quadrant events is strongly influenced by their local environment.

The structure envelopes are also shown in figures~\ref{fig:cond-v-xy-str}, \ref{fig:cond-v-yz-str} and \ref{fig:cond-v-xz-str}. For wall-attached Q2, wall-detached Q2 and wall-detached Q4 events, the envelope sizes of the causally enhanced structures are comparable to those of the causally suppressed structures. By contrast, for wall-attached Q4 events, the envelope of the causally enhanced structures is substantially larger than that of the causally suppressed structures. This is consistent with the feature analysis in figure~\ref{fig:geo-feature}, which shows that size is not a dominant feature for wall-attached Q2, wall-detached Q2 or wall-detached Q4 events, but is important for wall-attached Q4 events.
}

\bibliographystyle{jfm}
\bibliography{a-ref.bib}

\end{document}